# Kamodo: Simplifying Model Data Access and Utilization

Rebecca Ringuette[1,2], Lutz Rastaetter[2], Darren De Zeeuw[2,3], Asher Pembroke[4], and Oliver Gerland IV[4].

[1]ADNET Systems Inc., 6720B Rockledge Dr., Suite 504, Bethesda, MD, USA, 20817,
[2]The Community Coordinated Modeling Center, NASA Goddard Space Flight Center, Greenbelt, MD, USA, 20771,
[3]Catholic University of America, 620 Michigan Ave., N.E., Washington, DC, USA, 20064,
[4]Ensemble Government Services LLC, 4005 Buchanan St., Hyattsville, MD, USA, 20781.

*Abstract*

To address the lack of user-friendly software needed to simplify the utilization of model data across Heliophysics, the Community Coordinated Modeling Center (CCMC) at NASA's Goddard Space Flight Center has developed a model-agnostic method via Kamodo for users to easily access and utilize model data in their workflows. By abstracting away the broad range of file formats and the intricacies of interpolation on specialized grids, this approach significantly lowers the barrier to model data access and utilization for the community while adding exciting new capabilities to their tool boxes. This paper describes the direct interfaces to the model data, called model readers, and a basic introduction on how to use them. Additionally, we detail the planned approach for including custom interpolation codes, and include current progress on specialized visualization developments. The CCMC is maintaining Kamodo as an official NASA open-sourced software to enable and encourage community collaboration.

*Section 1: Introduction*

Currently, not many Heliophysics software resources currently provide the capability for users to utilize modeled data without a 'deep dive' into the particulars of model data format and data interpolation. Access to a small selection of models is often only available via custom codes produced in the literature, but the codes are not typically publicly available, are written in various programming languages, and often have non-uniform syntax. One notable example is the CCMC's library of visualization and analysis code. The CCMC maintains a large library of code in several languages that produces generalized interpolation and visualization capabilities available on the CCMC website and the applications available there. The code library includes some custom visualizations for specific models, similar to those presented in section 4. However, the library of code is not publicly available and portions of it depends on proprietary software licenses. These codes also do not provide direct functional access to the model data. Kameleon[1], a model access code developed previously at the CCMC, was an early attempt to solve these issues. However, the Kameleon software was primarily written in C and some Python, which made it cumbersome for some users to use and for some contributors to add support for new models and functionalities. Another example is SpacePy[2], an open-source package written in Python, which offers access to data output from the SWMF/BATSRUS model, but is not easily extensible to other model outputs (Morley et al. 2011, Toth et al. 2007). One exception is the Pysat[3] software package, which currently provides limited access to outputs from two models in the ionospheric domain, including a basic flythrough capability[4] (Stoneback et al. 2018). The CCMC team is collaborating with the pysat developers to extend their capability through a link to Kamodo, and have established interoperability with pysat and all of the core PyHC packages (Python in Heliophysics Community[5]: Barnum et al. 2022, Polson et al. 2022, Ringuette et al. 2022). Kamodo is now one of those core packages.

---
[1] https://ccmc.gsfc.nasa.gov/tools/kameleon/

To address this access and utilization challenge for model data, we are developing a model-agnostic method for users to interact with all model data hosted at CCMC using Kamodo, a unique Python package specifically designed for this challenge (Pembroke et al. 2022). The final product will enable simple access and utilization function calls for Heliophysics model data in several domains, including model output from the thermosphere, ionosphere, magnetosphere, heliosphere, and solar physics models hosted at CCMC. These capabilities will include direct interaction with model data via model readers, a satellite flythrough functionality, and additional software tools all designed to have identical (or nearly identical) function calls regardless of the model data chosen.

All capabilities discussed here are currently available through NASA's Kamodo GitHub repositories as open-source code[6]. Kamodo-core, the second link provides the core functionality of Kamodo, including function registration, composition, unit conversion, automated plotting, and remote procedure call (RPC) infrastructure. The first link hosts the model and data readers specific to space weather, the satellite flythrough functionality, customized visualization functions, and a variety of other functionalities tailored to meet the needs of CCMC users. That portion of Kamodo depends on the Kamodo-core library and subclasses the Kamodo class defined therein. Kamodo-core is available for installation via pip[7]. Installation of the entire Kamodo software requires additional steps which are described on the NASA github repository linked below (first link in footnote 6).

In this paper, we focus on the direct interface to the model data, called model readers, and their capabilities in comparison with the missing infrastructure for model data. As we expand the library of models, some of the model readers will require specialized interpolation and coordinate conversion codes. We therefore also describe our plan for incorporating such codes into the existing structure while adhering to our standard of model-agnostic syntax. Analysis of model data sets also vitally depends on the available visualization capabilities, which in turn require specialized development of advanced visualizations. We include several examples of advanced visualizations we have developed for this purpose. Although we will include a brief description of the model-agnostic flythrough and additional functionalities in the summary of this paper, we defer a more thorough consideration of these developing functions to a future work.

---

[2] https://spacepy.github.io/
[3] https://pysat.readthedocs.io/en/latest/index.html
[4] https://pysatmodels.readthedocs.io/en/latest/index.html
[5] https://heliopython.org/, https://ror.org/012prn105
[6] https://github.com/nasa/Kamodo and https://github.com/nasa/kamodo-core
[7] https://pypi.org/project/kamodo/

We begin by describing Kamodo in Section 2. This is followed by a tour of the model readers' usage and structure in Section 3, including a subsection dedicated to custom interpolations. Section 4 describes the standard visualization capabilities and displays several custom visualizations now available. We demonstrate the cross-model capability of the model readers in Section 5 by showing the simple syntax needed to easily plot and compare the same variable from more than one model. We conclude our work with a summary in Section 6.

*Section 2: What is Kamodo?*

Kamodo is an open-source software originally developed at the CCMC to specifically address the complex problem of simplifying access to and utilization of the Heliophysics model data hosted there. Although model data access and utilization is a complex problem in its own right and would easily justify its own software package, Kamodo was also designed with a broader array of applications in mind, such as data functionalization, unit conversions, function composition, automatically generated interactive plotting, and model-data comparisons. These capabilities were built in Python using SymPy[8] and Plotly[9] as described in Pembroke et al. (2022) (Python: Rossum & Boer 1991, SymPy: Meurer et al. 2017, Plotly: Plotly Technologies Inc. 2015). Due to its advantageous software design, Kamodo also has the flexibility to enable its use both within and outside of Heliophysics. Examples of these and other Kamodo core capabilities and exciting developments are given in Pembroke et al. (2022) and in the Kamodo core documentation[10].

The task of developing the stated functionality for all the model data output types hosted at CCMC is a daunting task (see Figure 1 for a sample of the models hosted at CCMC). Our development plan is to create model readers for a selection of models in each Heliophysics domain and develop various analysis functionalities (e.g. the afore-mentioned flythrough) based on the structure of those model readers. At the time of writing, we offer model readers for twelve Ionosphere-Thermosphere model outputs (teal category in Figure 1), two Magnetosphere model outputs (yellow category in the same figure), and a model output in the atmosphere-ionosphere domain (WACCM-X, not shown in Figure 1). Specifically, the library of model outputs supported in Kamodo includes the CTIPe (Coupled Thermosphere Ionosphere Plasmasphere Electrodynamics model, Codrescu et al. 2008), IRI (International Reference Ionosphere model, Bilitza 2018), GITM (Global Ionosphere Thermosphere Model, Ridley et al. 2006), OpenGGCM (global magnetosphere outputs only, Open Geospace General Circulation Model, Raeder et al. 2001), SWMF (ionosphere electrodynamics and magnetosphere outputs, Space Weather Modeling Framework, Toth et al. 2007), AMGeO (Assimilative Mapping of Geospace Observations, AMGeO Collaboration 2019), TIEGCM (Thermosphere Ionosphere Electrodynamics General Circulation Model, Qian et al. 2013), WACCM-X (Whole Atmosphere Community Climate Model with thermosphere and ionosphere extension, Liu et al. 2018), DTM (Drag Temperature Model, Bruinsma 2015), SuperDARN (Super Dual Auroral Radar Network, both output types, Cousins & Shepherd 2010), Weimer (Weimer 2005), ADELPHI (AMPERE-Derive ELectrodynamic Properties of the High-latitude Ionosphere model, Robinson et al. 2021), and the WAM-IPE model outputs (the coupled Whole Atmosphere Model and Ionosphere Plasmasphere Electrodynamics models, Maruyama et al. 2016 and Fang et al. 2022).

---

[8] https://www.sympy.org/en/index.html
[9] https://plotly.com/
[10] https://ensemblegovservices.github.io/kamodo-core/

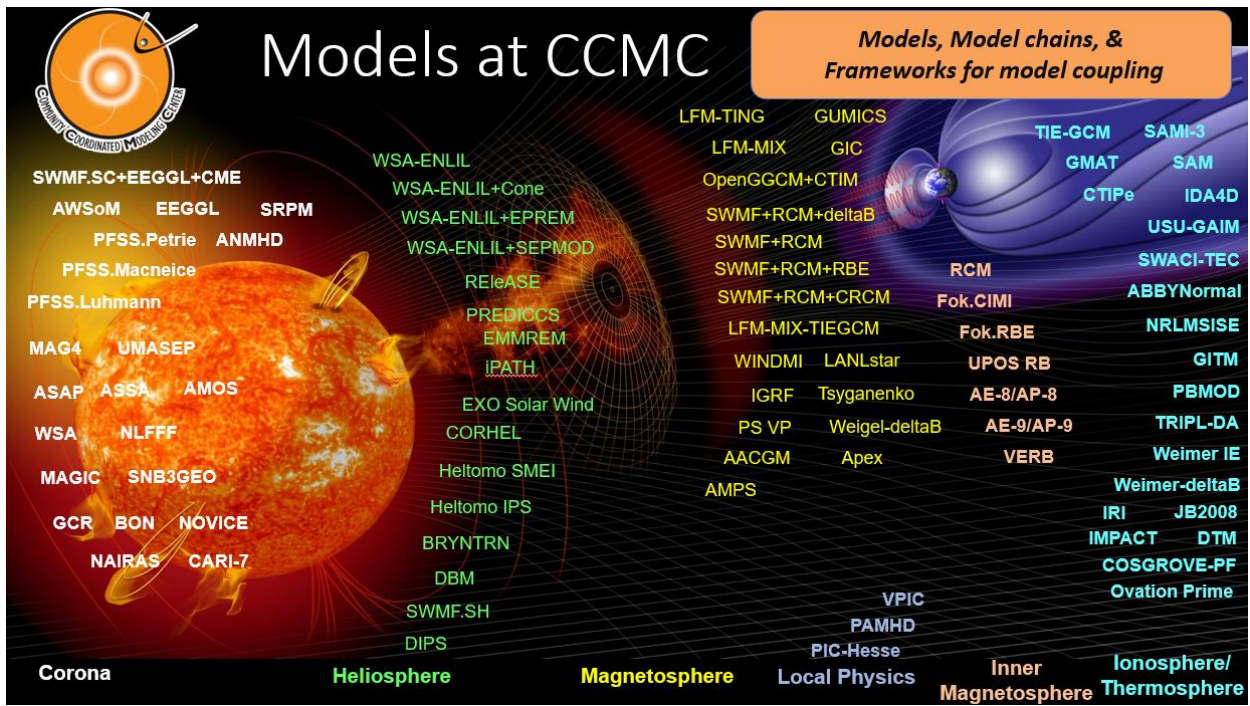

**Figure 1:** Models, model chains, and model frameworks hosted at CCMC. Model names (full or abbreviated) supplying model data in a particular domain are typed in the same color font as the domain name at the bottom of the figure.

A number of model readers are currently in development in a variety of domains. In the ionosphere realm, a model reader is in progress via collaboration for the SAMI3 model (Sami3 is A Model of the Ionosphere, Huba & Krall 2013). We have recently expanded our coverage of magnetosphere model outputs by adding the magnetosphere portion of the SWMF model outputs. This model reader is the first to include a custom interpolator script called from another language, and will be the example used for a number of collaborations in various domains (see section 3.3 and section 6 for details).

There are also readers in Kamodo for observational data such as datasets from CDAWeb (Coordinate Data Analysis Web[11]) and satellite trajectories from SSCWeb (Satellite Situation Center Web[12]). These readers facilitate data-model comparisons but are beyond the scope of this paper. By developing a library of model reader examples and the associated documentation, we are building the resources for collaborators to add additional models to Kamodo. Additionally, we are providing motivation for such collaborations by developing useful analysis tools these new model readers will plug in to. Such collaboration is greatly simplified by the Kamodo software being and remaining open-sourced, which was one of the prioritized requirements in its original development at the CCMC.

*Section 3: Model Readers*

We define the model readers in Kamodo as the direct model-agnostic interface between the user and the model data selected. As a standard, these model readers all provide access to the model data selected as functionalized variables using a simple, uniform command (Figure 2). This simple command abstracts away all the model-specific details from the user, especially the intricacies of data interpolation on complex grids and the complications of dealing with various file formats. Since these model readers are built using Kamodo, the user can then use the various features of Kamodo - such as automatic interactive plot generation (Figure 9), function composition, and unit conversions - in their analysis of the model data selected. Model readers are verified by comparing sample outputs from each model reader either to the previously verified corresponding outputs from the current version of the CCMC online visualization or in collaboration with model developers using sample outputs provided by the model developer. For a list of model outputs currently supported in CCMC's Kamodo, see Figure 3.

*Section 3.1: Using the Model Readers*

The model-agnostic command to functionalize the selected data and variables is shown on the third line of the grey box in Figure 2. As a standard, the only information required in the function call to a model reader is the complete file path to the model data. The necessary precursors to this command are given on the previous two lines and include an import statement for the model directory (and other functionalities) and a statement to retrieve the chosen model readers (second line). The third line functionalizes all of the variable data found in the chosen file directory (*file_dir*) – 19 variables in this case. As is common, the file path to the model output data is specific to the data structure on one's machine or online environment, so the model readers alone cannot provide this information. The user must do so. The value of the string used in the second line ('GITM') can be chosen from the output of the command shown in Figure 3.

---

[11] https://cdaweb.gsfc.nasa.gov/index.html/
[12] https://sscweb.gsfc.nasa.gov/

Once Kamodo is installed from the Github repository (first link in the footnote above), there are two methods to approach the pair of commands needed to retrieve the chosen model reader. The user can either obtain the name of the model reader by directly accessing the /kamodo_ccmc/readers/ directory, or by using the feature provided by the model_wrapper script in the /kamodo_ccmc/flythrough/ directory as shown in Figure 2. For example, the first two lines of code in Figure 2 would be replaced by the first two lines of code of Figure 4, which return the same model reader for the GITM model output as before, but with the additional work of finding the correct model reader name. We recommend the method shown in Figure 2 for simplicity.

```
import kamodo_ccmc.flythrough.model_wrapper as MW
reader = MW.Model_Reader('GITM')
kamodo_object = reader(file_dir)
kamodo_object
```

$$\rho_n(\vec{r}_{GDZsph4D})[\frac{kg}{m^3}] = \lambda(\vec{r}_{GDZsph4D})$$

$$\rho_{nijk}(time[hr], lon[deg], lat[deg], height[km])[\frac{kg}{m^3}] = \lambda(time, lon, lat, height)$$

$$T_n(\vec{r}_{GDZsph4D})[K] = \lambda(\vec{r}_{GDZsph4D})$$

$$T_{nijk}(time[hr], lon[deg], lat[deg], height[km])[K] = \lambda(time, lon, lat, height)$$

**Figure 2:** Example of the simplicity of the command for a model reader. The first two lines retrieve the model reader instance for the chosen model, and the third line functionalizes all of the variables found in the model data located in the given file directory *file_dir*. Calling the 'kamodo_object' variable as on the last line of code prints the LaTeX output representing each variable functionalized from the model data given, as shown in the output box below the code (2 of 19 functionalized variables shown). Note two functionalized versions of each variable are included by default (see documentation[13]).

For many science questions, the user needs to know the time and coordinate range of the data and the variables present in a given file directory. We have a few simple functions in the model_wrapper script referenced above to easily provide this information. Given the model name and the file path where the desired model output is stored, the user can simply execute the line of code in Figure 5 to be presented with the beginning and end times (in UTC) of the files. The function also returns the same two times as datetime[14] objects for other uses. To retrieve a list of variables present in the chosen data, users can execute the single line of code in Figure 6. Such a functionality is important since the variables included in a given data set is often a subset of those accessible through a given model. The output contains the name of each variable followed by a list of metadata, including a description, an integer, the coordinate system acronym, the coordinate system type, a list of coordinate names, and the units. The same function can be used to search the library of models for a given variable (e.g. temperature) even without any model data. The single command in Figure 7 can be executed after *kamodo_object* is created in Figure 2 (or Figure 4) for the desired list of variables to retrieve the minimum and maximum of the coordinates for each variable. The functionalities of the remaining keywords for these functions and each reader are demonstrated in the model reader tutorial notebook located in the notebook directory on NASA's Github repository.

---

[13] https://github.com/nasa/Kamodo
[14] https://docs.python.org/3/library/datetime.html

```
1  import kamodo_ccmc.flythrough.model_wrapper as MW
2  MW.Choose_Model()
```

```
Possible models are:
ADELPHI: AMPERE-Derived ELectrodynamic Properties of the High-latitude Ionosphere https://doi.org/10.
1029/2020SW002677
AMGeO: Assimilative Mapping of Geospace Observations https://doi.org/10.5281/zenodo.3564914
CTIPe: Coupled Thermosphere Ionosphere Plasmasphere Electrodynamics Model https://doi.org/10.1029/200
7SW000364
DTM: The Drag Temperature Model https://doi.org/10.1051/swsc/2015001
GAMERA_GM: Grid Agnostic MHD for Extended Research Applications - Global Magnetosphere outputs http
s://doi.org/10.3847/1538-4365/ab3a4c (coming soon)
GITM: Global Ionosphere Thermosphere Model https://doi.org/10.1016/j.jastp.2006.01.008
IRI: International Reference Ionosphere Model https://doi.org/10.5194/ars-16-1-2018
OpenGGCM_GM: The Open Geospace General Circulation Model - Global Magnetosphere outputs only https://
doi.org/10.1023/A:1014228230714
SuperDARN_uni: SuperDARN uniform grid output https://doi.org/10.1029/2010JA016017
SuperDARN_equ: SuperDARN equal area grid output https://doi.org/10.1029/2010JA016017
SWMF_IE: Space Weather Modeling Framework - Ionosphere and Electrodynamics outputs https://doi.org/1
0.1029/2006SW000272
SWMF_GM: Space Weather Modeling Framework - Global Magnetosphere outputs https://doi.org/10.1029/2006
SW000272
TIEGCM: Thermosphere Ionosphere Electrodynamics General Circulation Model https://doi.org/10.1029/201
2GM001297
WACCMX: Whole Atmosphere Community Climate Model With Thermosphere and Ionosphere Extension https://d
oi.org/10.1002/2017MS001232
WAMIPE: The coupled Whole Atmosphere Model - Ionosphere Plasmasphere Model https://doi.org/10.1002/20
15GL067312 and https://doi.org/10.1029/2022SW003193
Weimer: Weimer Ionosphere model https://doi.org/10.1029/2005JA011270
```

**Figure 3:** The library of models currently supported by CCMC's Kamodo. Following an import statement, the *Choose_Model* function can be used to show the library of models currently support in CCMC's Kamodo. The string at the beginning of each line is the string needed to retrieve the model reader script for the indicated model (used as the input in line 2 of the previous figure). Following this string is the full name of the model and the primary DOI(s) for that model.

Once the model reader object has been created, users may access the interpolator for each functionalized variable through similarly simple syntax using the features available through the core Kamodo package. Figure 8 demonstrates the various methods for calling a functionalized variable's interpolator. The first and third python commands show two equivalent methods to retrieve the LaTeX representation of a sample functionalized variable, either by using attribute notation (first line) or dictionary notation (third line). In this case, the two functionalized variables are two versions of the same variable: a non-gridded interpolator (top two input/output pairs) and a gridded interpolator (bottom two input/output pairs). A non-gridded interpolator requires as input a list of position lists (specifying positions in a 1-dimensional arrangement) as shown on the second line for two positions. A gridded interpolator accepts one or more coordinate positions to extract a lower dimensional slice or a hyperslab as shown on the fourth line for the same position. Note the input coordinates - time in hours since midnight of the first data file in the file directory, longitude in degrees, latitude in degrees, and altitude in km - and the resulting values for each are identical. Additional examples are included in the Kamodo core documentation. The strings representing the variable names in Figures 7 and 8 are model-specific and can be acquired using a variety of methods demonstrated in the various notebooks on the NASA Github repository (see footnote 8) and in Figure 6, including executing the command *kamodo_object.detail()*.

```python
from kamodo_ccmc.readers.gitm_4Dcdf import MODEL
reader = MODEL()
kamodo_object = reader(file_dir)
kamodo_object
```

$$\rho_n(\vec{r}_{GDZsph4D})[\frac{kg}{m^3}] = \lambda(\vec{r}_{GDZsph4D})$$

$$\rho_{nijk}(time[hr], lon[deg], lat[deg], height[km])[\frac{kg}{m^3}] = \lambda(time, lon, lat, height)$$

$$T_n(\vec{r}_{GDZsph4D})[K] = \lambda(\vec{r}_{GDZsph4D})$$

$$T_{nijk}(time[hr], lon[deg], lat[deg], height[km])[K] = \lambda(time, lon, lat, height)$$

**Figure 4:** Alternative method for loading a model reader. The first line of code imports the desired script. The second line loads the model reader class instance for the chosen model. The third and fourth lines and the output below are the same as Figure 2.

```python
MW.File_Times('GITM', file_dir)
```
```
UTC time ranges
-----------------------------------------
Start Date: 2015-03-17  Time: 00:00:00
End Date: 2015-03-19  Time: 01:48:59

(datetime.datetime(2015, 3, 17, 0, 0, tzinfo=datetime.timezone.utc),
 datetime.datetime(2015, 3, 19, 1, 48, 59, 996338, tzinfo=datetime.timezone.utc))
```

**Figure 5**: Retrieving the time range of a dataset. The UTC start and end date and time are printed for the chosen dataset located in *file_dir*. The returned values are the same dates and times represented as datetime objects.

Standard two-dimensional heat maps are easily attainable for any pair of dimensions through Kamodo. Once a multi-dimensional variable is functionalized with Kamodo, users can produce an interactive two-dimensional plot with a single line of code as demonstrated in Figure 9 below. The syntax simply requires the name of the desired variable (the gridded version), the names of the coordinates, and the desired values at which to slice the extra dimensions. The names of the variables can be retrieved via the method displayed in Figure 6. We note the longitude range varies with the coordinate system as defined by the SpacePy and AstroPy packages and refer the reader to the documentation included in the *Trajectory_Coords_Plots* notebook for the *ConvertCoord* flythrough function for more details (SpacePy: Morley et al. 2011, AstroPy: Astropy Collaboration 2018). Additional interactive plotting capabilities are demonstrated in section 4.

```
MW.Variable_Search('', 'GITM', file_dir)

The file directory contains the following standardized variable names:
--------------------------------------------------------------------------------
rho_N2 : ['mass density of molecular nitrogen', 13, 'GDZ', 'sph', ['time', 'lon', 'lat', 'height'], 'kg/m**3']
rho_N2plus : ['mass density of molecular nitrogen ion', 14, 'GDZ', 'sph', ['time', 'lon', 'lat', 'height'], 'kg/m**3']
rho_NO : ['mass density of nitric oxide', 20, 'GDZ', 'sph', ['time', 'lon', 'lat', 'height'], 'kg/m**3']
rho_NOplus : ['mass density of nitric oxide ion', 21, 'GDZ', 'sph', ['time', 'lon', 'lat', 'height'], 'kg/m**3']
rho_O2 : ['mass density of molecular oxygen', 22, 'GDZ', 'sph', ['time', 'lon', 'lat', 'height'], 'kg/m**3']
rho_O2plus : ['mass density of molecular oxygen ion', 24, 'GDZ', 'sph', ['time', 'lon', 'lat', 'height'], 'kg/m**3']
rho_O3P : ['mass density of atomic oxygen (3P state)', 27, 'GDZ', 'sph', ['time', 'lon', 'lat', 'height'], 'kg/m**3']
```

**Figure 6:** Retrieving the list and metadata for the variables found in the given dataset. The output contains the name of each variable followed by a list of metadata, including a description, an integer, the coordinate system acronym, the coordinate system type, a list of coordinate names, and the units. See text and documentation for more details, especially the START_HERE notebook in the */docs/notebook/* directory.

```
MW.Coord_Range(kamodo_object, ['rho_N2'])

The minimum and maximum values for each variable and coordinate are:
rho_N2:
time: [0.0, 49.81666564941406, 'hr']
lon: [-180.0, 180.0, 'deg']
lat: [-90.0, 90.0, 'deg']
height: [96.62001037597656, 640.3912963867188, 'km']
```

**Figure 7:** Retrieving the range for each coordinate for a given variable. The inputs here are taken from Figures 2 and 6. The minimum and maximum values for each coordinate the given variable depends on are given in a list next to the name of each coordinate, followed by the unit of that coordinate. For model readers developed at CCMC, time is always measured in hours from midnight UTC of the first day of the data in the given directory. The datetime representation of this date and time can be retrieved by executing the command *kamodo_object.filedate* once the chosen data has been functionalized. For the start and end times of a given dataset, see the *File_Times* command in Figure 5. Additional options and information for these functions can be found in documentation.

*Section 3.2: Model Reader Components*

To provide the described features and be compatible with higher level tools, we have determined the basic components necessary in the model readers and include a high-level description and demonstration here. Different approaches are necessary for different types of data to ensure fast execution times, as we will outline below. We refer the reader to the Kamodo documentation for more details (https://github.com/nasa/Kamodo).

```
kamodo_object.rho_n
```

$$\rho_n(\vec{r}_{GDZsph4D})[\tfrac{kg}{m^3}] = \lambda(\vec{r}_{GDZsph4D})$$

```
kamodo_object.rho_n([[5.5, 20., 55., 350.], [35.1, -45., -30., 200.]])
```

```
Time slice index 33 added from file.
Time slice index 34 added from file.
Time slice index 35 added from file.
Time slice index 211 added from file.
Time slice index 212 added from file.

array([5.38050868e-12, 2.95876514e-10])
```

```
kamodo_object['rho_n_ijk']
```

$$\rho_{nijk}(time[hr], lon[deg], lat[deg], height[km])[\tfrac{kg}{m^3}] = \lambda(time, lon, lat, height)$$

```
kamodo_object['rho_n_ijk'](5.5, 20., 55., 350.)
```

```
array(5.38050868e-12)
```

**Figure 8:** Example of various call methods to invoke the interpolator for a given functionalized variable. 'rho_n' is the functionalized neutral mass density from the GITM model reader. 'rho_n_ijk' is the gridded version of the same function. Note the input coordinates and the resulting interpolated values are identical. See text and documentation for details.

The most important feature of the model readers is the interpolation of each variable's data in the entirety of the time range associated with the model data provided, including between data files. For coordinate grids that do not change with time or location, we provide a standard interface to a linear interpolation scheme based on SciPy's *RegularGridInterpolator*[15] (Virtanen et al. 2020). SciPy's interpolators are typically written in C and are called via various Python commands built to ensure fast and accurate execution. The *RegularGridInterpolator* offers several advantages over similar options available through SciPy which motivated our choice, such as not requiring equally spaced grid points, enabling linear interpolation of data in any number of dimensions, and avoiding more computationally expensive approaches (e.g. using Delauney triangulation).

We find these advantages to be particularly suited to the interpolation requirements for the model data hosted at CCMC that is currently supported by Kamodo, and likely also well-suited for a larger range of model data (see section 3.3 for a discussion of more complex scenarios). In general, this default interpolator calculates the values for 40,000 3D (time+2D space) or 4D (time+3D space) positions in about half a second for an entire day of data regardless of the model reader, but varies depending on the sizes of the data arrays. Additionally, the execution time to generate a Kamodo object for a single variable is typically faster than half a second assuming all necessary file conversions have already been performed (see below). Although these execution and interpolation times are in most cases a bit slower than the current scripts used in the CCMC's online visualization services, the range of application is far greater and thus well worth the effort.

---

[15] https://docs.scipy.org/doc/scipy/reference/generated/scipy.interpolate.RegularGridInterpolator.html

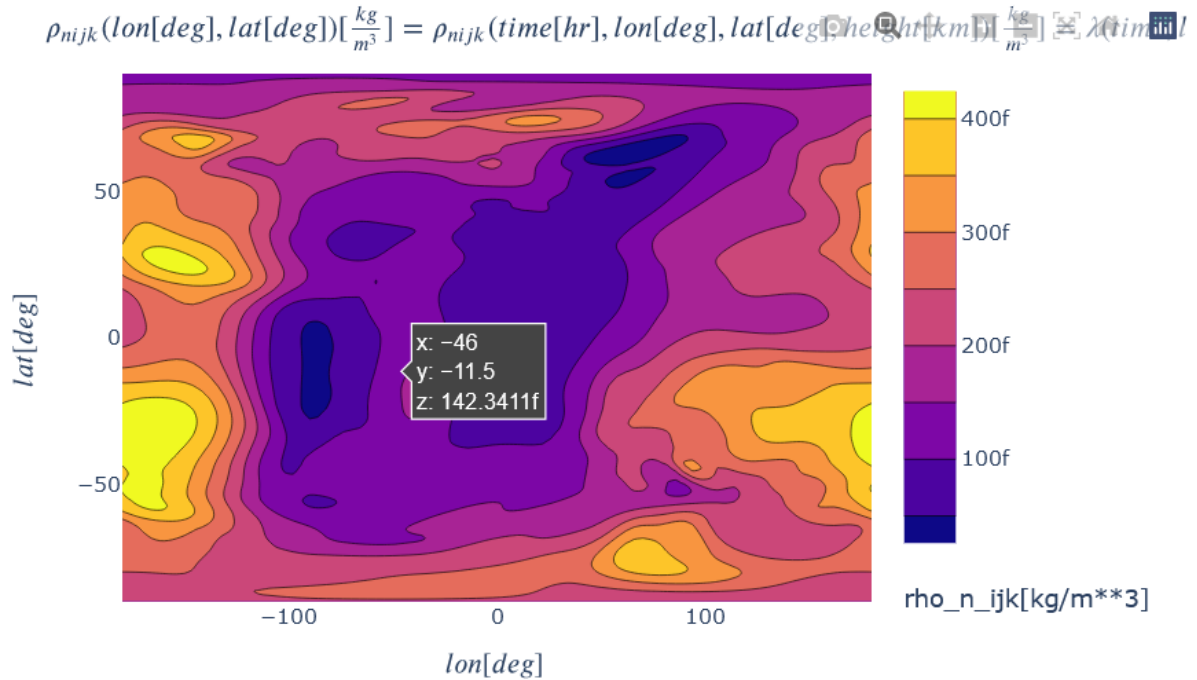

**Figure 9:** Example of the automatic interactive plot generation available through Kamodo. The grey box at top contains the command to produce a latitude vs. time plot of the neutral density variable ('rho_n_ijk') functionalized by the GITM model reader. The box in the plot at the mouse pointer's location (here with a grey background) displays the time, latitude, and neutral temperature values at the user's mouse position in the units given in the LaTeX representation of the function at the top of the plot. Note the symbols at the top right of the plot enable the typical Plotly interactive plotting features, such as zooming and saving. The 'time slice…' outputs below the command block are messages printed by the lazy interpolation scheme indicating which time slices were loaded to produce the plot. See text and documentation for more details.

Each model reader also provides customized access to the data format specific to each model. For model output data in a standard format such as netCDF4 files, this access is trivial to code, but the issue becomes decidedly more complex for other formats, especially for uncompressed or compressed binary files (e.g. compare the simplicity of the iri_4D.py model reader script for a netCDF4 model output file to the combined gitm_4D.py model reader and the gitm_tocdf.py file converter scripts for an uncompressed binary file). The file converter script for the OpenGGCM global magnetosphere model output is written in Python and deals with compressed binary files through FORTRAN-based functions via the f2py interface (Peterson 2009). This script will be called from the OpenGGCM model reader and is publicly available on Kamodo's Github[16].

The decision to functionalize the entire dataset stored in a given file directory has an important consequence on the interpolation requirements. We can not assume that the entire dataset can be read into memory all at once, with the exception of one-dimensional time series data (e.g. a variable depending only on time and not position). In the one-dimensional case (*interp_flag=0*), the entire dataset is read into memory from all of the relevant files and assigned the *interp1d* SciPy interpolator[17]. This option can also be used for higher dimensional variables not requiring lazy interpolation, in which case the standard *RegularGridInterpolator* interpolator is assigned (see documentation). For variables dependent on time, we have designed a system of lazy interpolators to read into memory and functionalize only the data corresponding to the time steps needed for the interpolation requested. This allows us to interpolate through datasets that are larger than a typical computer's memory (~16 GB).

An efficient implementation of lazy interpolation requires the software to know which data files hold which time values and variables before opening the file. We reduce the interpolation time by calculating and storing the mapping between the time values and the file names during the first execution of the chosen model reader on a given file directory. To calculate this mapping, the script searches for and opens every data file in the directory to retrieve the time value (or values) stored in each data file. The time values retrieved from each file are then saved in different formats in two text files saved to the directory. By separating the mapping into two files, we successfully manage the slightly more complex case of having multiple files per timestep where the data is split by variables (e.g. as opposed to splitting by processor number for the same variable). We avoid repeating this step for every model reader execution due to the additional time this process takes and the additional cost this incurs for executions in cloud environments, particularly when model data is stored in s3 buckets. The mapping between requested variables names and the relevant data files only requires one file of each type to be opened and queried, and so does not need this extra step. During the model reader execution, both mappings are assembled into memory and used to initialize the interpolation for each variable regardless of the interpolation method used for each variable.

Three lazy interpolation options are available based on the structure of the data files. The simplest lazy interpolation method (*interp_flag=1*) is designed for model outputs with the data separated into one or more files per timestep. The other two lazy interpolation methods have similar logic and are designed for model data outputs with more than one time step per data file (a.k.a. time chunks). The *interp_flag=2* option is for time-chunked data files that easily fit into a typical computer's memory (<16 GB RAM), while the *interp_flag=3* option is for larger time-chunked data files. For smaller files (option 1 or 2), the given time slice or chunk is loaded into memory as needed. For larger files (option 3), the data for a given time step selected from the time chunk is loaded into memory to reduce the demand on computer memory.

---

[16] https://github.com/nasa/Kamodo/tree/master/kamodo_ccmc/readers/OpenGGCM
[17] https://docs.scipy.org/doc/scipy/reference/generated/scipy.interpolate.interp1d.html

For each interpolation call, the software determines which time steps or chunks are required to perform the given interpolation, typically one time step on either side of the value or the corresponding time chunk for option 2. Once the model time values are chosen, the mapping between these time values and the files is used to determine which files holds the required data. This information is then sent to the model reader, which retrieves the requested data, perform any data wrangling needed, and returns the data to the interpolation program. For option 2, the time chunk containing the time value is returned with an additional time step added from the next time chunk if available to properly interpolate between the time chunks. Finally, the standard SciPy interpolator is created for each time step or time chunk and an *interp1d* interpolator is assigned for the time dimension, resulting in two layers of interpolation executions per interpolation call. We find this approach reduces the total interpolation execution time by avoiding repeated creations of an interpolator as more times are requested. We further reduce computation time for time values requested that are already present in the data by only using the data for that given time value (e.g. plot requests).

The logic for all of the interpolation methods assumes all of the data for a given variable and timestep are located in a single file. This approach is also used for model data where the data for different variables and the same timestep are stored in different files. In these and similar cases, the mapping between the variables and the set of files is captured in the model reader logic (e.g. the GITM model reader). However, the current structures do not easily account for model outputs where the data for a given time step and variable is distributed among many files (e.g. split by processor number). For these and other complex situations, each of the three lazy interpolation schemes have a built-in feature to easily incorporate a custom interpolator (see section 3.3) or custom read routines.

Additionally, execution speed becomes an issue when the model data is stored in text files or binary files as compared to CDF, netCDF, h5 files or similar formats. To navigate these issues, the model readers automatically convert each data file in the directory into a netCDF file the first time the model reader is called, but only if the original data is stored in a non-ideal file format. If the user adds data files to the same directory, then the user will need to delete the time files previously created by the model reader in the same directory and rerun the model reader to include the new files. This behavior is required to avoid executing a directory listing command each time the model reader is run, which is time-consuming and not cost-effective in a cloud environment.

The file conversion process also removes any data deemed extraneous by model developers, such as status logging variables, and performs any data reformatting and manipulation necessary for all of the remaining variables regardless of the variable names requested. We find this decision substantially reduces the model reader execution time on subsequent calls, and in some cases also appreciably reduces the total file size, excluding the original file sizes (Table 1). However, we currently choose to keep all original model output files to allow future reprocessing and to preserve data that is currently not included in the converted files.

Table 1 presents a comparison of execution times for the five model readers requiring file conversions. The initial execution times include the file conversion time and the time needed to functionalize a single variable (second column), while the final execution times only include the functionalization time for the same single variable. The given uncertainty represents the standard deviation calculated using five execution times in identical conditions. The values reported for the GITM (Ridley et al. 2006), Weimer, ADELPHI, and SWMF ionosphere electrodynamics model outputs are for one day of data, while the values reported for the OpenGGCM (global magnetosphere) (Raeder et al. 2001) model output are for one hour of data. Note the final execution times are significantly shorter than the initial execution times for all model readers presented.

**Table 1:** Comparison of execution times for first and subsequent calls to selected model readers.

| Model Reader | Initial Execution Time (s) | Final Execution Time (s) | Initial Total File Size (GB) | Final Total File Size (GB) |
|---|---|---|---|---|
| GITM | 974. ± 19. | 0.563 ± 0.033 | 20.3 | 8.37 |
| SWMF (IE) | 200.7 ± 1.4 | 0.301 ± 0.029 | 3.89 | 1.97 |
| ADELPHI | 21.0 ± 1.0 | 0.309 ± 0.059 | 0.131 | 0.00783 |
| Weimer | 84.73 ± 0.73 | 0.296 ± 0.016 | 1.92 | 0.157 |
| OpenGGCM (GM) | 736 ± 59[†] | 0.107 ± 0.022 | 4.38 | 15.6 |

[†]The initial execution times do not include the execution time of the OpenGGCM (GM) model reader. The larger final file size results from decompressing the original compressed binary file format.

Many ITM (Ionosphere-Thermosphere-Mesosphere) model outputs in spherical coordinate systems include variables that depend on pressure level instead of height or radius. The pressure level coordinate grid is defined differently in each model output and varies both spatially and temporally, complicating the interpolation of a given satellite position or a requested variable plot at a constant height. However, the pressure level coordinate grid values are defined as constant reference points in the model outputs, and so invite the use of the typical *RegularGridInterpolator*, albeit in an innovative way. Due to the presence of this relationship in multiple models, we have created a model-agnostic method based on the standard interpolator to invert each model's relationship between pressure level and height[18] without any further assumptions on the underlying data.

Each model output including variables dependent on pressure level also gives height as a function of time, longitude, latitude, and pressure level. This relationship must be inverted to return the pressure level at a given time, longitude, latitude, and height. We accomplish this with a few simple steps. First, the height variable is functionalized and gridded to enable slicing. For each unique trio of time, longitude, and latitude values, we interpolate the model-provided pressure level coordinate grid values through the height function, producing a set of height values corresponding to each pressure level grid value at that time and longitude-latitude location. We then create a one-dimensional interpolator using the *interp1d* interpolator provided by SciPy with the interpolated height values as the 'X' values and the model-provided pressure level values as the 'Y' values. Once created, the 1D interpolator returns the pressure level for a requested height. This provides a custom inverted relationship between the height and pressure level for each time, longitude, and latitude combination requested. This allows the model readers to automatically redefine each variable dependent on pressure level as a new variable dependent on height via Kamodo's function composition capability using only the model data provided. The generalization of this code to work with all relevant models provided the framework to efficiently include custom interpolators (see section 3.3).

The chosen structure of the model readers results in a linearly increasing execution time for an increasing number of variables. Figure 10 presents the execution times of each model reader in the CCMC Kamodo model library plotted against the number of variables functionalized. Both the regular and gridded versions of each variable were requested in these calls (*gridded_int = True*, see documentation), and any necessary file conversions took place beforehand. For variables using a lazy interpolation scheme (with two or more dimensions), model reader executions do not load any model output data into memory except for the coordinate grids. The variables functionalized for these timing tests were therefore limited to be at least two dimensions or more. Unfortunately, this restriction resulted in the SuperDARN tests being restricted to three possible variables (white squares and triangles). The DTM model output has only nine variables total, and so lacks a data point for the tenth variable (purple triangles). The Weimer model output is excluded since it only has one or two variables, depending on the dataset.

In addition to the lines and symbols for each model reader, the average execution time is also plotted as a black dashed line with one sigma error bars. The average execution time for one variable is 0.50 ± 0.19 s, as shown in the inset of Figure 10, with an average increase of 0.4618 ± 0.0046 s of execution time per additional variable. The standard deviation of the execution times also increases with increasing number of variables from 0.19 s for one variable up to 0.94 s at 10 variables, with the increasing spread dominated by the WACCM-X execution times for higher numbers of variables. In general, the execution times of the model readers tend to be within one standard deviation of the average execution time per variable. Such close tracking is due to a distribution of properties among the readers, such as the number of unique coordinate grids, the size of each coordinate grid, the presence of model-specific coordinates requiring additional initialization steps (e.g. pressure level), the complexity of the variable-file mapping logic, and other similar considerations.

---

[18] See the *PLevelInterp* function in the reader_utilities.py script in the *kamodo_ccmc/readers/* directory on our GitHub (https://github.com/nasa/Kamodo).

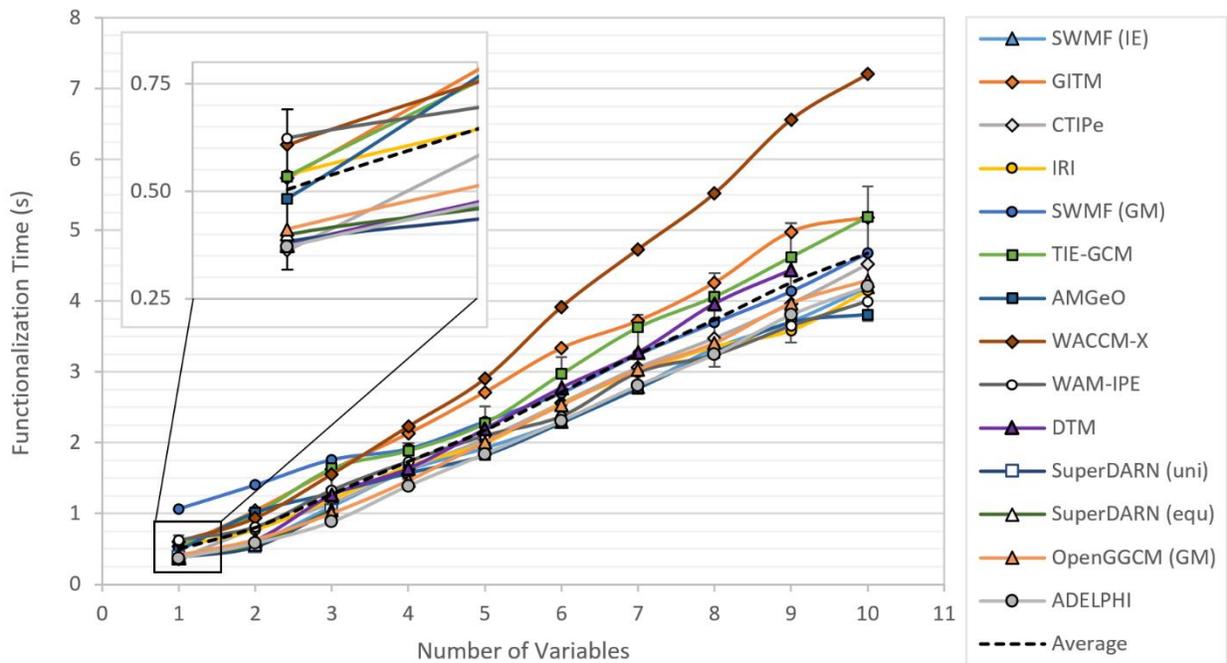

**Figure 10:** Total execution time for selected model readers vs number of requested variables. Colors and shapes with solid lines indicate model readers (see legend). The average execution time across the model readers sampled is plotted with a black dashed line. Error bars on the average line show the 1σ standard deviation at each variable number. The execution times are typically closely grouped with the exception of the SWMF (GM) execution time at low numbers of variables, and the WACCM-X execution time at high numbers of variables. The inset at top left shows the distribution of execution times using one variable. See text for more details.

There are, however, two exceptions to this trend. The SWMF (GM) model reader execution time (blue line with circles) is well above the error bars for small numbers of variables, and the WACCM-X model reader execution time (brown line with diamonds) is well above the error bars for four or more variables. The behavior of the SWMF (GM) execution time is likely due to the additional initialization steps required to properly prepare the custom interpolation C code for execution compared to the other codes. In contrast, the larger execution times of the WACCM-X model reader code at larger variable numbers is probably caused by the large size of the coordinate grids – roughly similar to OpenGGCM (GM) – in combination with the pressure level inversions automatically functionalized at larger variable numbers for the chosen set of variables.

Additional features included in the model readers are automatic linear interpolation in time between model data output files, scalar and vector averaging at the poles (for model outputs on spherical grids that do not include the poles, see documentation), mapping between the often cryptic model variable names and standardized variable names, and alignment of the model data with a standard coordinate system defined by the SpacePy and AstroPy packages (SpacePy: Morley et al. 2011, AstroPy: Astropy Collaboration 2018). Although model runs are typically generated with a continuous time range, the outputs are produced by collecting all data for a set number of timesteps or a set time range into a file, generating an artificial time gap between the files corresponding to the time resolution of the model run. Interpolation across this time gap is necessary to properly represent the full data set produced by the model run. The simplest approach would be to use nearest neighbor interpolation for time values between files, where the time value in the neighboring file closest to the requested time value would be chosen, and the typical linear interpolation would be performed for the spatial components. However, this approach is not consistent with the linear interpolation for all time and spatial components at all other times. So, we have chosen to implement linear interpolation in time between data files as a standard in all model readers. Interpolation between files is accounted for in the lazy interpolation method described above. We refer the interested reader to the Kamodo onboarding instructions document on the NASA Kamodo Github repository for more details.

Model outputs on spherical grids do not always extend to include the poles. In most cases, this is due to the grid values reported from the model being located in the center of a grid cell that includes the poles (e.g. a grid cell located at 89.5 degrees latitude with one degree resolution). In these situations, we have implemented additional logic to calculate the expected value of the variable data at the pole. For scalar quantities such as density or cartesian components, this logic simply averages the range of values next to the pole in question and populates the array vector corresponding to the pole's latitude with that value. This is done separately for each time value. For spherical components, particularly the non-radial components, we take the vector average of the same values, also separately for each time, and populate the corresponding values at the pole with those values. Again, we refer the interested reader to the onboarding document for more details and examples of this logic.

Variable name mapping is typically determined in coordination with model developers by comparing each variable description in the model output to other similar variables already implemented for other models. This mapping is necessary to determine the best LaTeX representation of each variable to best correlate with the standard representation of that variable in literature (e.g. 'rho' for total density or 'rho_n' for neutral density). Each model reader stores the variable mapping in the *model_varnames* dictionary as key, value pairs accessible through the methods described above (e.g *print(key, value[0] for key, value in model_varnames.items()* after importing the *model_varnames* dictionary from the desired reader). Similarly, alignment of the variable data with a standard coordinate system defined by the SpacePy and AstroPy packages is also typically done in coordination with model developers.

Other features not described here ensure compatibility with higher level functionalities, such as the flythrough function. For further details and additional examples, we refer the reader to the documentation, example notebooks, and code hosted at the NASA Kamodo GitHub repository[19].

*Section 3.3: Incorporating Custom Interpolators*

The standard interpolator, SciPy's RegularGridInterpolator, and the interpolation schemes described in the last section are only appropriate for spatial grids that are static in time and in each dimension, orthogonal, and do not include refinement (nested regions of successively finer resolution around objects). For models using non-standard grids, particularly in the geospace domain, custom interpolation routines are required, which are often written in languages other than Python for efficiency. This section describes our approach to easily accommodate those interpolators, using our work on the SuperDARN (an ITM model) and SWMF/BATS-R-US magnetosphere output data as examples.

The SuperDARN model outputs are generated with two types of grid resolutions: a default grid of uniform resolution and an equal area grid. The default grid is trivial to interpolate across using the methods previously described, but the equal area grid has a longitude grid that changes resolution with latitude. In this situation, a custom interpolator was required. The custom interpolator function was simple enough to write in Python using the *interp1d* SciPy interpolator, but required the creation of a keyword option in the interpolation schemes to accept a custom interpolator (the *func_default* keyword, see Figure 11 below).

Functionalizing the SuperDARN equal area output served as a test for the inclusion of a custom interpolator, but did not include a cross-language bridge since the custom interpolator was written in Python. We chose to functionalize the SWMF/BATS-R-US magnetosphere output data next because the output currently lacks a four-dimensional (spatial+time) interpolator for its complex grid, it is a commonly used model in magnetospheric physics, and required initializing and calling a custom interpolation routine written in C. Also, this work will be easily transferable to other model outputs utilizing an octree block grid structure, including the GUMICS (Grand Unified Magnetosphere-Ionosphere Coupling Simulation; Janhunen et al. 2012) and ARMS (Adaptively Refined Magnetohydrodynamics Solver; Wyper et al. 2018) models. More generally, the method we determine to efficiently incorporate this work into a Kamodo model reader will

---

[19] https://github.com/nasa/Kamodo

Page 18 of 37

be extended to other custom interpolators, including those we are planning via collaborations (see section 6).

The SWMF contains many component models, including the Block-Adaptive Tree Solar wind Roe Upwind Scheme (BAT-S-RUS: Powell et al. 1998) MHD solver that uses a block adaptive grid to resolve fine structures and shocks within the modeling domain (Figure 12). Each block is a cube composed of a fixed amount (at least 4x4x4) of same-sized cells. Blocks are arranged to fill the simulation domain at the coarsest refinement level. Blocks are refined into 8 'children' by splitting each cell into two in each of the 3 dimensions. This forms an octree structure that can be followed to determine the leaf block (one without any children) for any given position in the simulation domain.

```python
def func(i):
    '''i is the file number.'''
    # get data from file
    file = self.pattern_files[key][i]
    cdf_data = Dataset(file)
    data = array(cdf_data.variables[gvar])
    cdf_data.close()
    return data

# define and register the interpolators
self = RU.Functionalize_Dataset(self, coord_dict, varname,
                                self.variables[varname],
                                gridded_int, coord_str,
                                interp_flag=1, func=func)
```
A

```python
def func(i):  # i is the file/time number
    # get data from file(s)
    cdf_data = Dataset(self.pattern_files[key][i])
    lat = array(cdf_data.variables['lat'])
    lat_keys = [str(lat_val).replace('.', '_').replace('-', 'n')
                if lat_val < 0 else 'p'+str(lat_val).replace(
                    '.', '_') for lat_val in lat]
    lon_dict = {lat_val: array(cdf_data.groups[lat_key]['Lon'])
                for lat_key, lat_val in zip(lat_keys, lat)}
    data = {lat_val: array(cdf_data.groups[lat_key][gvar])
            for lat_key, lat_val in zip(lat_keys, lat)}
    cdf_data.close()

    # assign custom interpolator
    interp = custom_interp(lon_dict, lat, data)
    return interp

# functionalize the variable dataset
tmp = self.variables[varname]
tmp['data'] = zeros((2, 2, 2))  # saves execution time
self = RU.Functionalize_Dataset(
    self, coord_dict, varname, tmp, gridded_int, coord_str,
    interp_flag=1, func=func, func_default='custom')
```
B

**Figure 11:** Comparison of a simple lazy interpolation call (A) with a custom lazy interpolation call (B). Panel A is from the GITM model reader and simply reads the data for the required time step from the file and returns the data. Panel B, from the SuperDARN equal area model reader, also reads in the data for the time step from the file, but must navigate a more complex data structure (a grouped h5 file) and call an external custom interpolator routine designed for that data structure. Note the return statement of the function *func* in Panel A returns a NumPy[20] array, while the corresponding statement in Panel B returns an interpolator. This different behavior is communicated to the interpolation software by setting the *func_default* keyword to 'custom' at the bottom of Panel B. See documentation and code for more details.

The SpacePy package contains readers for many of the SWMF's component models, including the pyBats package to read outputs of BATS-R-US. Kamodo utilizes the *SpacePy.PyBats.IdlFile* class and the *read_idl_bin* function to read the outputs formatted for visualization by the IDL scripts, including the MHD quantities of number density, pressure, velocity, current density and magnetic field on cell centers. PyBats, however, does not include an interpolator that takes advantage of the octree block structure of the BATS-R-RUS output[21].

---
[20] https://numpy.org/

To produce the standard visualization services at the CCMC, we have been using a custom package of functions written in C that traverses the octree grid structure to identify the leaf block for a given position and then implement a trilinear interpolation of the MHD variables at the position, similar to the RegularGridInterpolator in 3D. Python is an interpreted language and is much slower than compiled libraries derived from C sources when performing searches and numerical computation involving large arrays of data, so we have implemented a cross-language approach to call the C programs and have linked a shared library of the C functions into Kamodo.

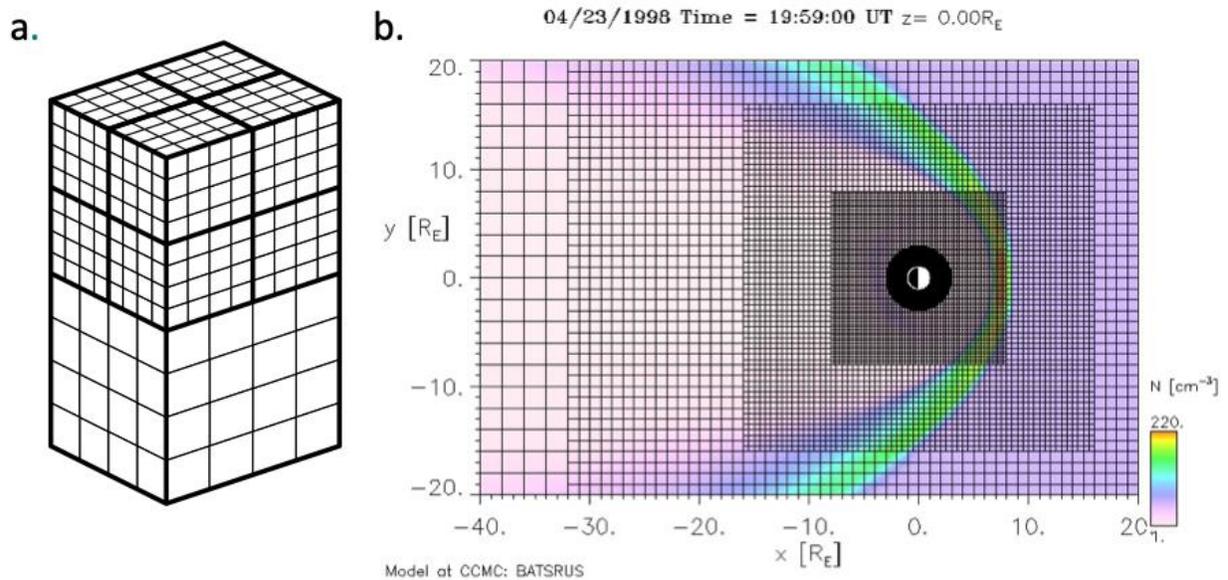

**Figure 12:** BATS-R-US grid structure: (a) Blocks forming the octree grid (from Fig. 1 of Toth et al. 2006) used by the SWMF BATS-R-US model's magnetosphere (cartesian as shown), solar corona and heliosphere (spherical coordinates). (b) The CCMC's online visualization service's rendition of plasma (H+) number density N (colors) and the grid (black), sliced at Z = 0 R_E (Earth radius), in a CMCC magnetosphere simulation. Solar wind flows from the right and forms the magnetosheath (green to red colors) around the magnetosphere (lighter colors on left) and the Earth (shown at origin, surrounded by a black region excluded from the simulation domain). The sun is to the right in the figure.

---

[21] We are collaborating with the SpacePy developers to add a new keyword to the PyBats.IdlFile() script to prevent the reader of binary output files from automatically rearranging the grid positions close to each other by physical location rather than keeping them arranged by block identity and logical cell position (i,j,k) within each block. A pull request has been submitted for PyBats.IdlFile() to add the keyword *sort_unstructured_data* to allow the user to sort data as was done in the binary reader before or leave the data unsorted (see https://github.com/spacepy/spacepy/pull/584 for more information).

To be compatible with many SWMF/BATSRUS versions, the tree structure is reconstructed using only the positions contained in the original formatted binary output file and disregarding the tree information being written to a separate file in newer model versions. The interpolator implements a search in the octree grid structure and then performs trilinear interpolations within the leaf block to arrive at the desired interpolated value. In principle this can be re-implemented in Python by defining the tree search and then using SciPy's RegularGridInterpolator instances defined for each block for each quantity. However, this will become unmanageable and inefficient with the number of blocks exceeding 100,000 for large grids, and each block requiring the definition of a separate interpolator object.

Using the cffi package in Python[22], one can easily specify interface functions among the C functions in the library that Python needs to access. Existing source files and other functions defined within the build script are packaged into a shared library that is then loaded like any other Python package (e.g. line 32 of the script referenced in the footnote below). This functionality has been successfully tested on both Windows and Mac machines and available in the current CCMC Kamodo version[23], as demonstrated in the Kamodo plot of the functionalized $B_z$ variable (Rastaetter et al. 2022 and Figure 13 below). This work will be extended to other model outputs with similar grids upon completion (e.g. GUMICS and ARMS), and more generally to other models requiring custom interpolation routines. The same method described here for including custom interpolation scripts can also be applied for custom file reader and conversion routines (e.g. the file converter for OpenGGCM compressed binary files mentioned earlier).

*Section 4: Custom Visualizations*

Large datasets are typically investigated beginning with visualization methods, often resulting in complex scripts written by each user for each dataset. Kamodo greatly simplifies this process by offering a wide range of publication quality interactive graphics with a single line of code. Although any visualization package can be implemented in the Kamodo-core package, Plotly was chosen as the default visualization package during the early stages of development due to the quality and quantity of features available at the time. Plotly is a free open-source data visualization toolkit in Python that enables customizable images in a wide variety of styles. The visualizations are highly interactive (zoom, pan, probe, rotate), can incorporate time sliders, and are easily embeddable into web pages or dashboards, while maintaining their interactive functionality.

---

[22] https://cffi.readthedocs.io/
[23] The link below is to the Python script that uses FFI from the cffi module (available via pypi) in Python to make a library. The cdef() method defines C data types and function declarations that Python needs to know about. The set_source() method defines C source code (lines starting with // are comments) needed in addition to code in files listed in the sources[] list. External libraries needed are listed in libraries[] (here: the math library libm.so is added with the -l compilation flag). The compile() method builds the shared library file (e.g., _interpolate_amrdata.cpython-37m-darwin.so on MacOSTM using Python 3.7).
https://github.com/nasa/Kamodo/blob/master/kamodo_ccmc/readers/OCTREE_BLOCK_GRID/interpolate_amrdata_extension_build.py

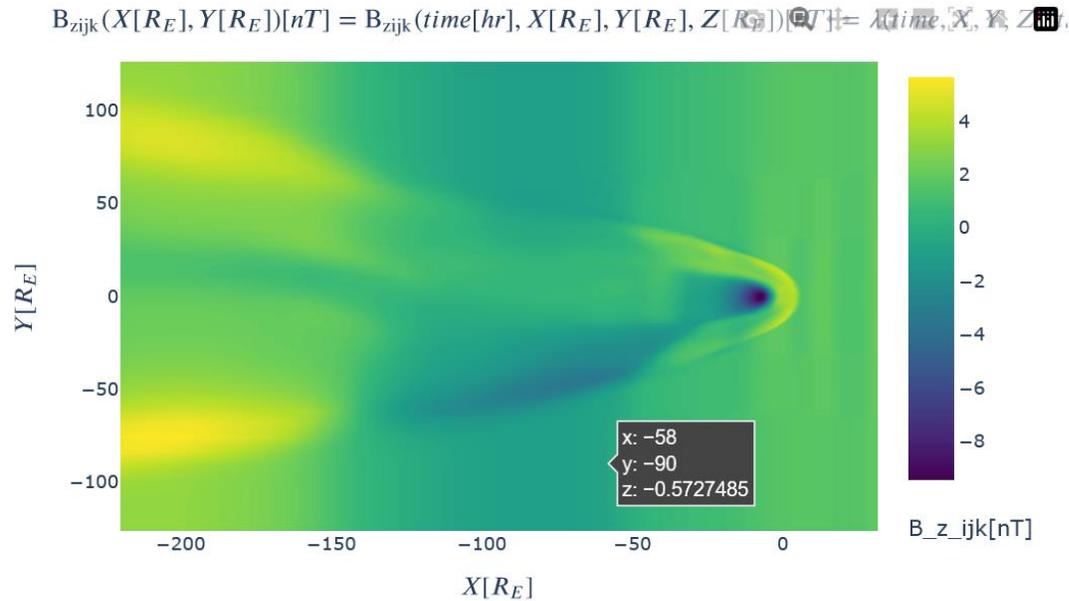

**Figure 13:** Kamodo visualization of the gridded B_z variable in the SWMF magnetosphere output. The LaTeX representation of the $B_z$ variable is given at the top and is overlaid by the standard PlotLy buttons on the top right. The interactive visualization shown displays the z component of the magnetic field at a Z value of 15.7 $R_E$ at 1.2 hours after midnight UTC on Dec 18, 2010 for the *TreShunda_James_071322_1* model run[24]. The custom interpolator described in the text was used to interpolate the displayed values at the indicated time and Z position. The grey box shows the X and Y coordinates of the mouse on the plot and the $B_z$ value at the same location. Axes are shown in $R_E$ and colors correspond to the colorbar at right. The bow shock of the Earth's magnetosphere reacting to the solar wind is evident on the right side of the plot. The solar wind originates from the Sun to the right (not shown), and the dark roughly circular area contains the Earth. The plot is produced using the same command as in Figure 9, but with a few extra lines to change the color map and the plotting method. The 'Time slice…' messages are produced by the underlying lazy interpolation logic. See documentation and text for details.

Kamodo builds upon the Plotly capabilities to provide users a library of automatically generated plots. This internal registry of plotting functions is indexed by the function's calling argument shape and return output shape. For example, if a plot is requested for a function called with N points returning N points, then a 1D line plot will be given (e.g. Figure 14). See the table of options in Kamodo's visualization documentation and the associated figures for a complete list of the default plot types for a variety of function signatures[25].

---

[24] The data used to produce this plot is available at
https://ccmc.gsfc.nasa.gov/RoR_WWW/output_files/KAMODO_DEMO/SWMF-GM/TreShunda_James_071322_1/

We include a few examples to demonstrate the application of a selection of the default plotting types to functionalized science data using Kamodo. Kamodo registers the data, functionalizes it with user provided interpolation, and displays interactive plots via a simple command based on the shape of the data. Figure 14 shows a simple example with an array of 265 timestamps passed to a function that returns the synthetic density at each time and includes the commands necessary to generate the plot. The 'kamodo.plot' function call returns an object that can be displayed, exported in several ways, or further modified or combined to make other figures. As an example of combining Plotly objects, Figure 15 shows a plot object returned in Kamodo (the 'ror' variable) from a model output (a CCMC Run-On-Request product[26]) and another returned from a data source (GOES-12 data, obtained from the HAPI (Weigel et al. 2021) interface to CDAWeb) (GOES: Geostationary Operational Environmental Satellites, Menzel et al. 1994). With a few lines of Python code in a Jupyter notebook[27] (Kluyver et al. 2016), the plots can be combined, the labeling updated, and the resulting figure displayed (see Figures 21 and 22 for an alternate method). Other options available through Plotly are described in Plotly's documentation[28].

The default Kamodo plots can also be used for datasets with higher dimensionalities through Kamodo's *partial_plot* keyword as shown in Figure 9. In a single line, the four-dimensional dataset is reduced to two dimensions and displayed in an interactive plot similar to those in documentation. This is done by interpolating the data to the given slice values in the two dimensions not plotted and using the original coordinate grid for the remaining two dimensions (see documentation for more details[29]). However, these default capabilities are not sufficient for the community to fully understand the data presented in model outputs, especially given the high dimensionality of those outputs.

While automated plots are important to enable users of all skill levels to get a meaningful plot easily, Kamodo and Plotly are capable of much more. A case study using the ability of Kamodo to extract data from simulation output along the path of a synthetic satellite trajectory will be used to highlight several custom plot options in Kamodo. (See Ringuette et al. 2022 for a thorough description of this and related capabilities.) For this example, a synthetic satellite trajectory was used to extract data from the CTIPe model run at the CCMC. One output from this extraction is the neutral density called 'rho' along the satellite path. Fitting into our

---

[25] https://ensemblegovservices.github.io/kamodo-core/notebooks/Visualization/
[26] https://ccmc.gsfc.nasa.gov/tools/runs-on-request/
[27] https://docs.jupyter.org/en/latest/
[28] https://plotly.com/python-api-reference/
[29] https://ensemblegovservices.github.io/kamodo-core/API/#multi-function-plots

```python
from kamodo import Kamodo, kamodofy
import numpy as np
import pandas as pd
from plotly.offline import iplot
```

```python
t_N = pd.date_range('Nov 9, 2018', 'Nov 20, 2018', freq = 'H')

@kamodofy(units = 'kg/m**3')
def rho_N(t_N = t_N):
    dt_days = (t_N - t_N[0]).total_seconds()/(24*3600)
    return 1+np.sin(dt_days) + .1*np.random.random(len(dt_days))

kamodo = Kamodo(rho_N = rho_N, verbose = False)
```

```python
fig = kamodo.plot('rho_N')
iplot(fig)
```

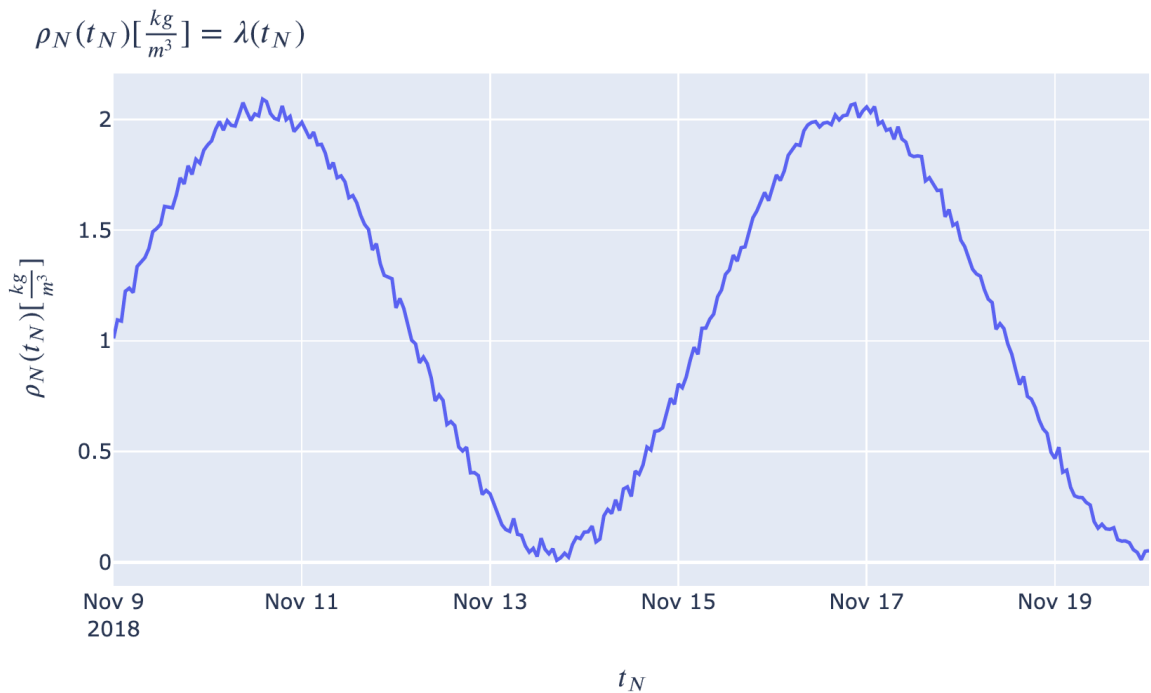

**Figure 14:** Python code to generate a 1D line plot from an input and output shape of N. The first block of code shows the import statements, the second block demonstrates the functionalization of a sample one-dimensional data set, and the third block shows the simple command that generates the plot at bottom. Note the LaTeX representation at the top left of the plot showing the functionalized variable, and the automatically labeled axes. The plot is fully interactive with zoom, pan, and similar features (not shown).

automated plot mapping, the function requires an (N,3) dimensional input (position) resulting in an (N) dimensional output (density) with time as extra information. This would normally result in a standard image showing a 3D heat map, but we can customize this in several ways. The position components can be broken out and shown in a variety of 1D plots easily, but since Kamodo also provides coordinate transformation by utilizing an extension of SpacePy and AstroPy, more complicated custom plots can be created. The same data can be shown as a 2D contour plot with position mapped to longitude and latitude in spherical - GEI (Geocentric Equatorial Inertial) coordinates as seen in Figure 16. We refer the reader to the SSCWeb for an explanation of the coordinate systems used here[30].

```python
fig = ror.get_plot(type = "1Dvar", var = "B_z")
fig2 = hapi2.get_plot(type = "1Dvar", var = "B_GSE_c")
fig.add_trace(fig2.data[2])
fig.update_yaxes(title_text='nT')
fig['data'][0]['name'] = 'B_z (MODEL)'
fig['data'][1]['name'] = 'B_z (DATA)'
iplot(fig)
```

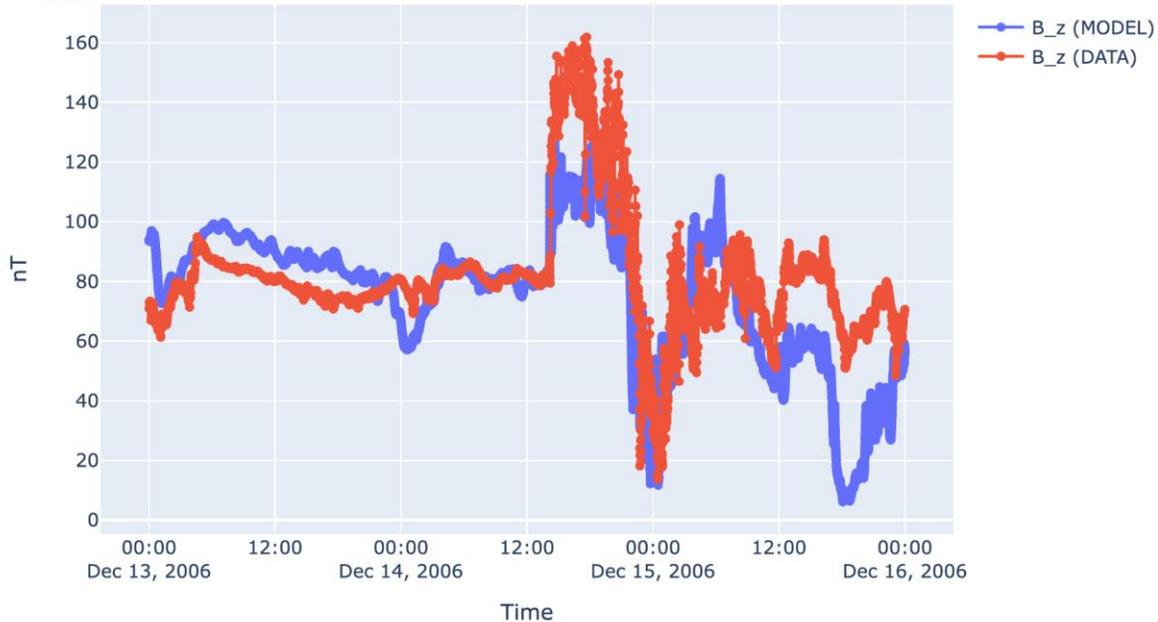

**Figure 15:** Python code to combine two Kamodo plot objects (top) and display the updated figure (bottom). The 'fig2.data[2]' selects the B_z value out of the three components of the B vector in that plot object. The data used to generate the plot was extracted from run GEM_CEDAR_082015_3, which can be obtained from the CCMC's Run-on-Request service[31].

The Python code above the figure illustrates the syntax of this plotting routine. In the code block displayed, two import statements are executed, then the *SFcsv_reader* function is used to read data from a csv file into a nested dictionary, and the *SatPlot4D* function is used to create and display a fully interactive plot, each accomplished in a single command. The various components of the 'cdf_dict' variable are the one-dimensional time series arrays containing the timestamps ('utc_time'), position information ('c1', 'c2', and 'c3'), and the variable calculation results ('rho') with the units. The three values from the metadata section of the dictionary used in the command are the name of the model used and the coordinate system information. Users can choose from a range of options such as the plotting coordinate system, labeling, trajectory slicing methods (e.g. 'all' in Figures 16 and 17, and 'orbitE' in Figure 18), and the option to save the plot to an html file with complete interactivity preserved in the file. The '2D' choice for the 'type' keyword produces the plot shown in Figure 16. Changing the value of this keyword produces other representations of the same data such as shown in Figures 17 ('2DPN') and Figure 18 ('3D'). See documentation for further information.

---

[30] https://sscweb.gsfc.nasa.gov/users_guide/Appendix_C.html
[31] https://ccmc.gsfc.nasa.gov/results/viewrun.php?domain=GM&runnumber=GEM_CEDAR_082015_3

```
from kamodo_ccmc.flythrough.wrapper_output import SFcsv_reader
from kamodo_ccmc.flythrough.plots import SatPlot4D
cdf_dict = SFcsv_reader('FakeFlightExample_CTIPe20s.csv')
var = 'rho'  # variable name
SatPlot4D(var, cdf_dict['utc_time']['data'], cdf_dict['c1']['data'],
          cdf_dict['c2']['data'], cdf_dict['c3']['data'],
          cdf_dict[var]['data'], cdf_dict[var]['units'],
          'GDZ', 'sph', 'GEI', 'all', 'CTIPe', type = '2D',
          htmlfile = 'Figure15.html')
-saving full html file:  Figure15.html
```

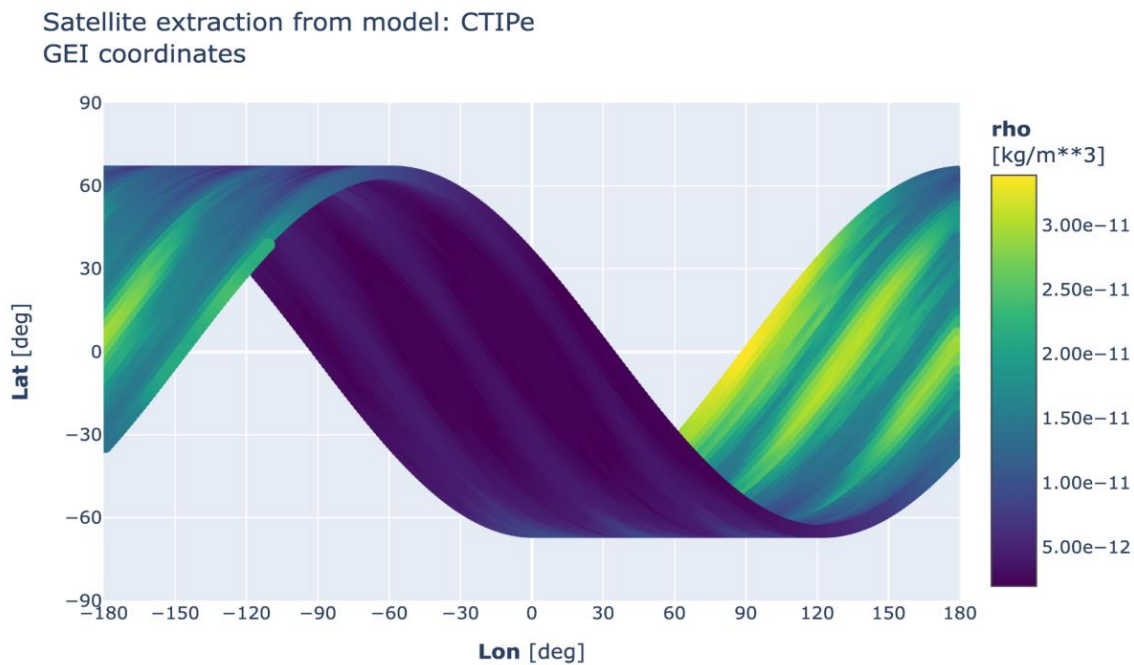

**Figure 16:** Mapping of full trajectory to latitude and longitude position colored by density. The code in the block above the plot shows the required syntax to produce the plot. Data is read in from a file with a single command, and then fed into the plotting command to produce the fully interactive plot shown at the bottom. The position of the satellite is shown in GEI spherical coordinates after an automatic conversion from the GDZ spherical coordinate system. The fully interactive plot is displayed in the notebook as well as saved to a html file, 'Figure15.html', which can be viewed later or placed on a web page.

Another view frequently used by scientists is a view from above the North pole. Figure 17 shows the density values and satellite trajectory over the Northern hemisphere in GEO (Geographic) coordinates with Earth continent outlines for context. This image also shows an information box hovering where the mouse pointer is positioned on the plot. Plotly images are interactive, showing additional meta-data for points near the current cursor location. In this case, the GEO X, Y, and Z positions, the GDZ (Geodetic) longitude, latitude, and altitude (lon, lat, alt) position, the density value and units, and the timestamp in UTC are shown.

Note that while these images are shown as static, the actual plots can be manipulated via zoom to view specific regions, and via rotation for 3D imagery. These plots were created using the same Kamodo plotting utility function with just a few minor changes to a couple of calling arguments. While they show the diversity of ways to display data and bring in contextual imagery, they still do not address the dynamics of time. To incorporate this dimension more interactively, we use a slider, directly supported by the Plotly features, to scroll through time. It can also be configured with play/pause buttons to start the animation. A sample capture of this type of display is shown in Figure 18, and only requires small changes to the plotting commands used in Figures 16 and 17.

These visualizations are, of course, highly customizable at a fairly novice level. In a Jupyter notebook or Python script, Kamodo can return an object that can be displayed or further modified before display. In the example in Figure 14 above, an object called 'fig' is returned and then displayed. This object can be modified through simple Python calls described through comprehensive Plotly online documentation or through direct object manipulation.

Two examples of advanced customization are given in Figures 19 and 20. Figure 19 was customized from a Kamodo default plot, and Figure 20 was built by layering in some advanced customization of magnetic vectors into a SatPlot4D default figure. Figure 19 shows a highly customized plot with the GRACE-1 satellite[32] (Wahr et al. 2004) flown through a GITM run at the CCMC (trajectory obtained from the SSCWeb). The model and satellite were chosen as a pair due to their overlapping altitudes. GITM simulates the ionosphere thermosphere system around the Earth from about 90 km to 600 km altitude, and the GRACE-1 satellite orbits at around 400 km to 500 km altitude. The satellite positions (retrieved from SSCWeb) were extracted from the GITM output with eight additional extractions made on a +/- 100 km box around the actual satellite position. The minimum and maximum values are shown as an envelope of values to represent a positional error bar. The satellite positions are also shown on the same plot using a separate Y axis scale (right axis). Probing the plot results in a display of all values at a given time.

---

[32] https://www.nasa.gov/mission_pages/Grace/index.html

```
var = 'rho'  # variable name
SatPlot4D(var, cdf_dict['utc_time']['data'], cdf_dict['c1']['data'],
          cdf_dict['c2']['data'], cdf_dict['c3']['data'],
          cdf_dict[var]['data'], cdf_dict[var]['units'],
          'GDZ', 'sph', 'GEO', 'all', 'CTIPe', type = '2DPN',
          body = 'lines', zoom = False,
          htmlfile = 'Figure16.html')
```

-saving full html file:  Figure16.html

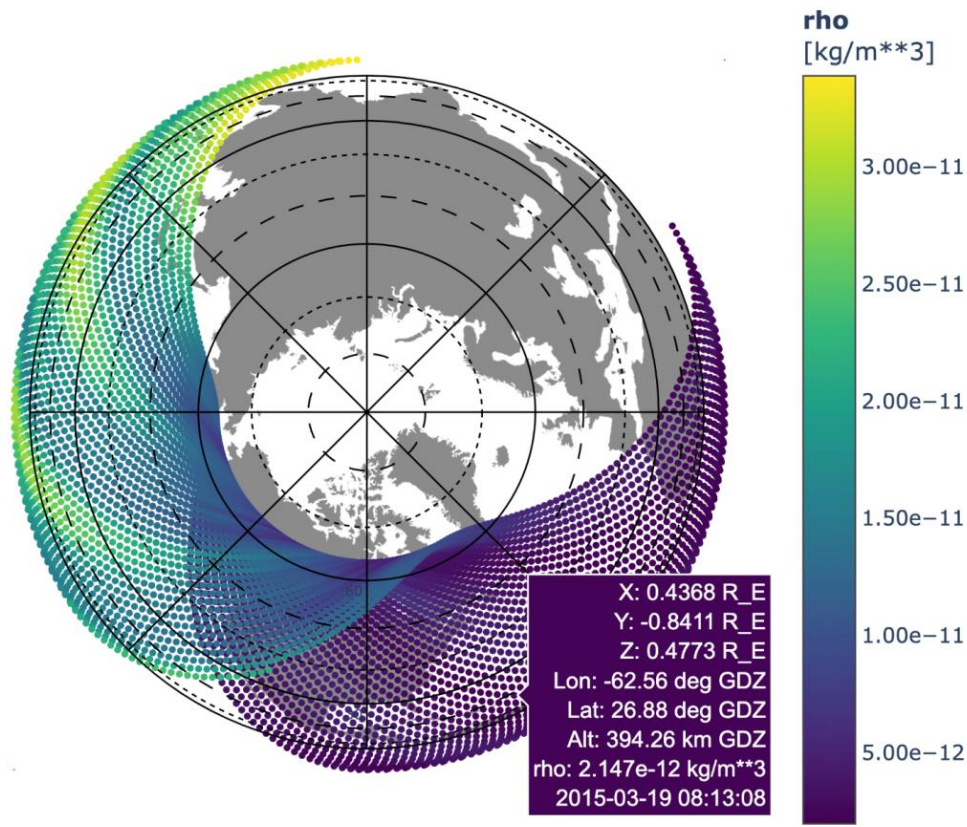

**Figure 17:** Position and density value displayed from a North polar view with additional hover meta-data. The code shown above the plot uses the same data retrieved in Figure 16. The plotting command (using *SatPlot4D*) is almost identical to that in Figure 16, with only a few altered keyword values to produce the plot shown. The position of the satellite is shown in GEO coordinates to allow a land/water layer under the plot for position context.

One other highly beneficial option provided with these Plotly visualizations is the ability to save them to html, allowing them to be used as dynamic plots in web browsers. The resulting file includes all the data necessary to reproduce the graphic and interact with it on the client side. To see the interactive nature of these, you can view Figures 16, 17, and 18, online[33].

```
var = 'rho'  # variable name
SatPlot4D(var, cdf_dict['utc_time']['data'], cdf_dict['c1']['data'],
          cdf_dict['c2']['data'], cdf_dict['c3']['data'],
          cdf_dict[var]['data'], cdf_dict[var]['units'],
          'GDZ', 'sph', 'GSE', 'orbitE', 'CTIPe', type = '3D',
          body = 'none', htmlfile = 'Figure17.html')
```

-saving full html file: Figure17.html

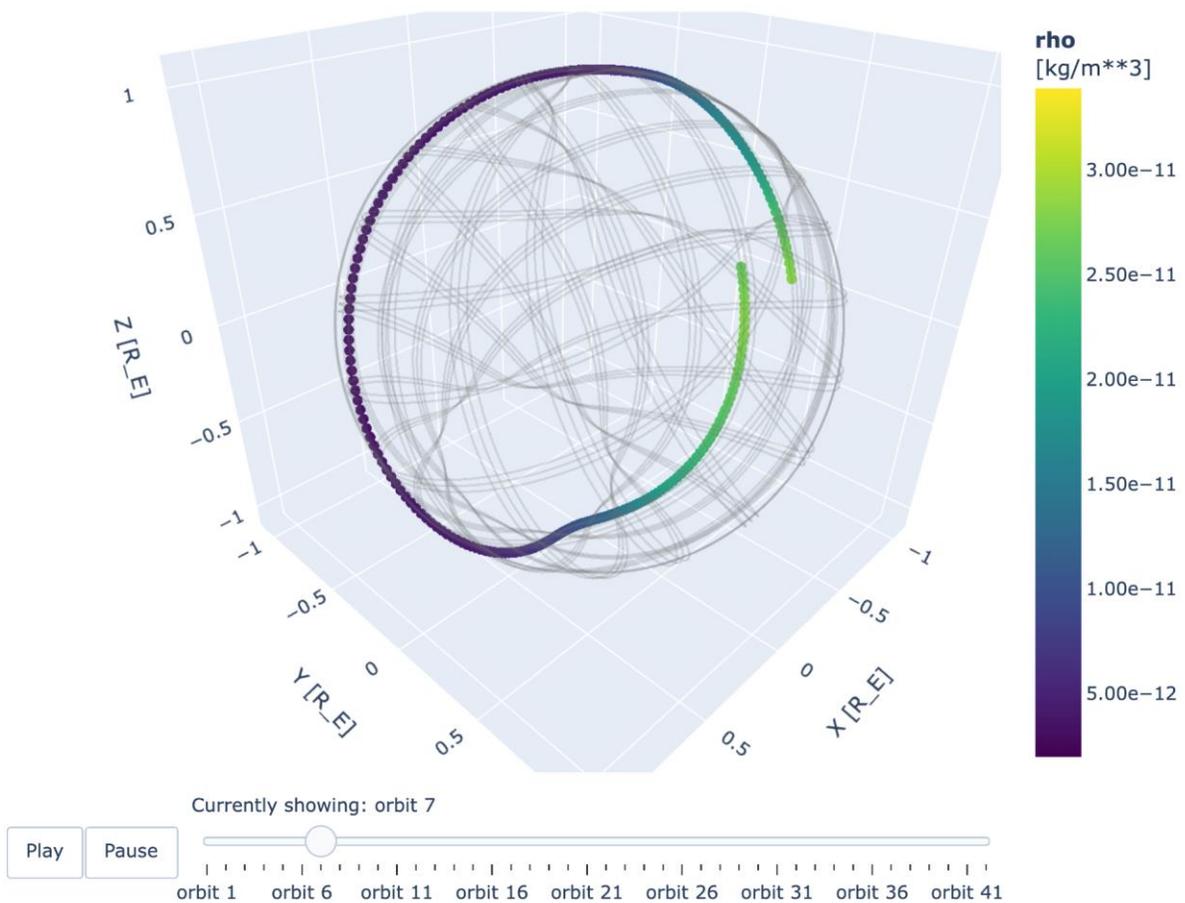

**Figure 18:** Full satellite trajectory broken up in orbits when the satellite crosses the equator heading North in GSE coordinates. The animation can be played or viewed by dragging the orbit slider on the bottom. Note the plotting command included above the figure uses the same data retrieved in Figure 16 and has almost identical values.

---

[33] https://ccmc.gsfc.nasa.gov/Kamodo/Figures/

Additional customized figures are also possible for the vectors in the magnetosphere and soon for other domains. Figure 20 shows magnetic field values calculated from model output for the GOES-13 satellite (a geosynchronous orbit around the Earth) (Hillger and Schmit 2009). The positions of the GOES-13 satellite were used to extract values from a global magnetosphere run at the CCMC. This extraction can then be visualized in Kamodo in several coordinate systems. The magnetic field vectors can also be added or removed and their cadence adjusted.

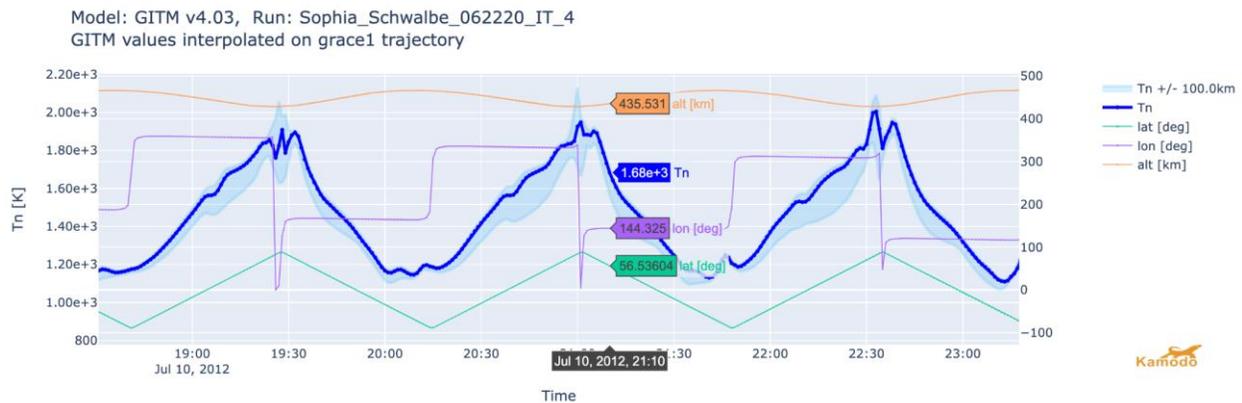

**Figure 19:** Customized plot with overlay of multiple line plots for an extraction of neutral temperature (Tn) values along the GRACE-1 satellite trajectory through a GITM model run. The dark blue line shows the neutral temperature (left axis), the soft blue shading shows the positional error determined from the additional extractions described in the text, and the remaining three lines show the position of the satellite in geodetic coordinates (latitude, longitude, and altitude on the right axis). The colored boxes contain the color-coded values of each variable on the plot, and update with the user's mouse movements. The data used in this plot is available through the CCMC's Run-on-Request service[34].

*Section 5: Comparing Data Across Models*

Combining the capabilities discussed so far results in a powerfully simple method to directly compare results across models, despite various file formats, disparate coordinate grids, and custom interpolation methods. Figure 21 demonstrates how this can be done with minimal coding using ion temperature modeled with the CTIPe and GITM models for a given storm day as an example. With a simple import statement, the ion temperature data from each model is functionalized with a few lines of code in the first two blocks, each block identical in syntax to the other. The functions are then collected into a single Kamodo object in the third block. This Kamodo object is used to plot the two datasets at identical slices in time and height for direct visual comparison in Figure 22. The same single line of code shown in Figure 9 is used to produce the two plots, each then followed by a line of code demonstrating some simple modifications for a smoother visualization. This example workflow can be adapted to plot and compare any variable from any model already included in Kamodo via a model reader.

---

[34] https://ccmc.gsfc.nasa.gov/results/viewrun.php?domain=IT&runnumber=Sophia_Schwalbe_062220_IT_4

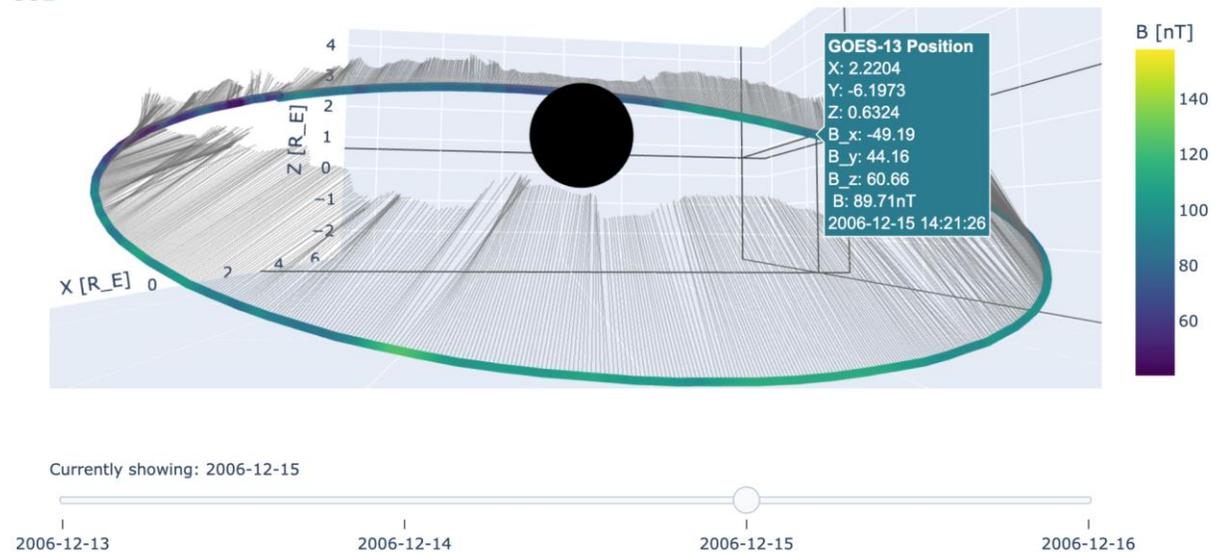

**Figure 20:** A customized plot showing the values from the GOES-13 satellite position extracted from a CCMC global magnetosphere run. The plot is broken up with a slider to select one day of data at a time. The color is the magnitude of the magnetic field at that position. The black spikes are the magnetic field vector at every 10th point along the satellite path. The mouse hover box shows the precise values at that position. The data used to generate the plot is the same dataset as in Figure 15.

*Section 6: Summary*

The model reader structure described above enables several important functionalities for the Heliophysics community. For the first time, users with little programming experience now have a simple, intuitive, and model-agnostic method to access and utilize model data from multiple physics-based and empirical models in their research without needing to understand the complexities of the various file formats and the interpolations necessarily involved. Basing this software functionality on Kamodo further simplifies analyses and comparisons of these data among multiple model outputs via the additional capabilities available through that package. Another important benefit of using Kamodo is the simplification of interoperability with other python packages (see Polson et al. 2022 and Ringuette et al. 2022 for two examples). We note our work is easily extensible to observational and model data in other fields. We refer the interested reader to the examples on the NASA Kamodo GitHub repository, especially the Kamodo onboarding instructions document, to achieve these same capabilities for their datasets. Users can obtain Heliophysics model data for a large collection of models by visiting the CCMC website[35].

---

[35] https://ccmc.gsfc.nasa.gov/

```
import kamodo_ccmc.flythrough.model_wrapper as MW
reader = MW.Model_Reader('CTIPe')
kamodo_object_ctipe = reader(ctipe_file_dir, variables_requested=['T_i'])
kamodo_object_ctipe
```
**A**

$$T_i(\vec{r}_{GDZsph4D})[K] = \lambda(\vec{r}_{GDZsph4D})$$
$$T_{iijk}(time[hr], lon[deg], lat[deg], height[km])[K] = \lambda(time, lon, lat, height)$$

```
# Retrieve model reader and access model data for the GITM model.
# Note the identical syntax!
reader = MW.Model_Reader('GITM')
kamodo_object_gitm = reader(gitm_file_dir, variables_requested=['T_i'])
kamodo_object_gitm
```
**B**

$$T_i(\vec{r}_{GDZsph4D})[K] = \lambda(\vec{r}_{GDZsph4D})$$
$$T_{iijk}(time[hr], lon[deg], lat[deg], height[km])[K] = \lambda(time, lon, lat, height)$$

```
# Collect functions into a single Kamodo object.
# Note the names of the coordinates each variable depends on.
from kamodo import Kamodo
kamodo_object = Kamodo()
kamodo_object['CTIPeT_i[K]'] = kamodo_object_ctipe['T_i_ijk']
kamodo_object['GITMT_i[K]'] = kamodo_object_gitm['T_i_ijk']
kamodo_object
```
**C**

$$CTIPeT_i(time[hr], lon[deg], lat[deg], height[km])[K] = \lambda(time, lon, lat, height)$$
$$GITMT_i(time[hr], lon[deg], lat[deg], height[km])[K] = \lambda(time, lon, lat, height)$$

**Figure 21:** First part of an example workflow directly comparing simulation results. The figure demonstrates the syntax to functionalize simulated data from two different models. In this case, simulated ion temperature data from the CTIPe and GITM models are functionalized with identical syntax (Panels A and B) and then combined into the same *kamodo_object* variable (Panel C). The two datasets are plotted in the next figure. This workflow can be adapted to be used for any variables and any models included in Kamodo via a model reader.

Several model readers for the magnetosphere domain are pending completion of the SWMF (global magnetosphere portion, Toth et al. 2007) model reader, and others are independently in development. The custom interpolator design for the SWMF/BATS-R-US model reader will be applied to model readers in development for the GUMICS and ARMS model outputs which have similar model data grid structures. The ARMS model reader will be our first venture into the solar corona domain. The model reader structure and custom lazy interpolation design will also be more generally applied to models developed via collaborations, such as GAMERA (Grid Agnostic MHD for Extended Research Applications, Zhang et al. 2019), VERB (Versatile Electron Radiation Belt code; Wang et al. 2019), MARBLE (Magnetosphere Aurora Reconnection Boundary Layer Explorer, Bard & Dorelli 2021), and RAM-SCB (Ring Current Atmosphere Interactions Model with Self-Consistent B field; Zaharia et al. 2010). Initial development has also begun for CIMI (Comprehensive Inner-Magnetosphere Ionosphere model, Fok et al. 2014), the first for drift-kinetic model output.

```
fig = kamodo_object.plot('CTIPeT_i', plot_partial={'CTIPeT_i': {'time': 12., 'height': 500.}})
fig.update_traces(colorscale="Viridis", ncontours=200,
                  contours=dict(coloring="fill",showlines=False))
fig
```

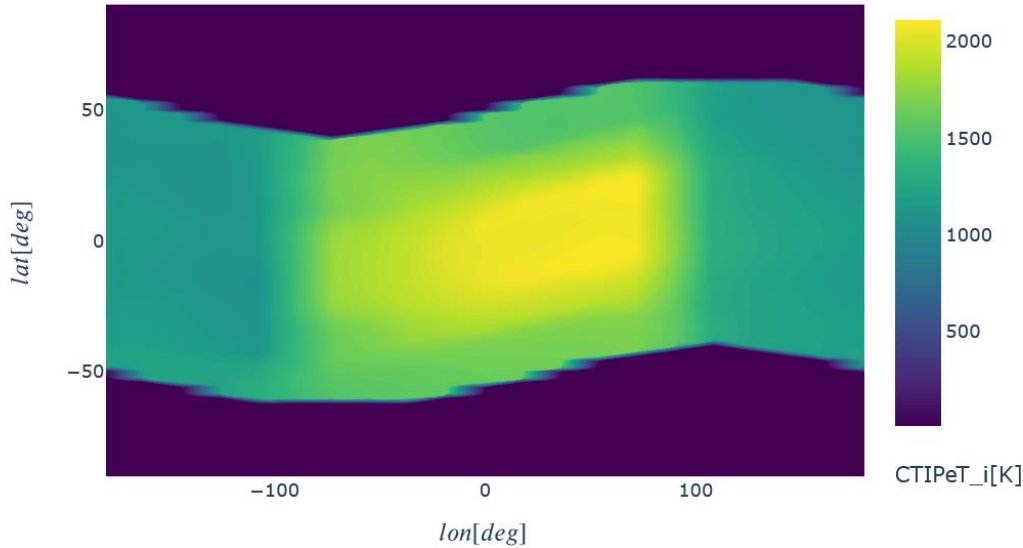

```
fig = kamodo_object.plot('GITMT_i', plot_partial={'GITMT_i': {'time': 12., 'height': 500.}})
fig.update_traces(colorscale="Viridis", ncontours=200,
                  contours=dict(coloring="fill",showlines=False))
fig
```

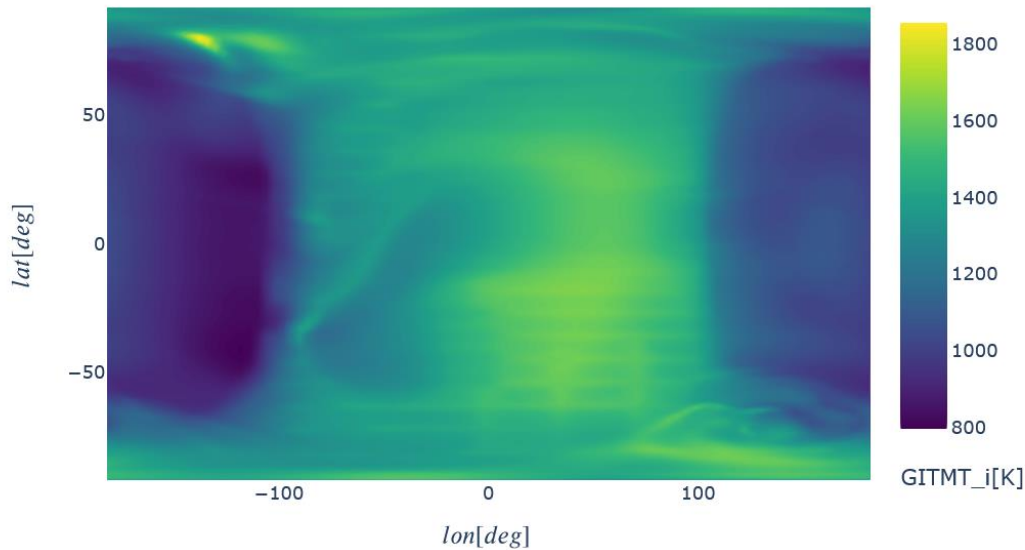

**Figure 22:** Second part of an example workflow directly comparing simulation results. In this portion, we use identical syntax to produce plots of simulation ion temperature data from the CTIPe (top) and GITM (bottom) models for the same day. The lazy interpolation messages are only printed the first time a given time slice is loaded, so are not shown here. This workflow can be adapted to be used for any variables and any models included in Kamodo via a model reader.

In parallel to our efforts to expand the library of models and domains represented in Kamodo, we are also working to expand the utility of Kamodo, both for general users and for CCMC applications. The unique model-agnostic capabilities possible through the model readers are currently the basis of development of further functionalities, such as a collection of similarly simplistic flythrough functions to 'fly' a given trajectory through a set of model data, which greatly simplifies analyses involving comparisons of observational and model data, and tools enabling studies of satellite constellation trajectories and arrangements (to be described in future works). Sample workflows for a variety of science applications are being added to the NASA Kamodo Github as they are completed, and are freely available for users to adapt to their own purposes (or contribute their own). These workflows are not only based on Kamodo's capabilities, but also include other Python software packages commonly used by the community (e.g. Polson et al. 2022).

At the CCMC, initial work has begun to use Kamodo for CCMC's Instant Run interface for a selection of models, and a sample usage for CCMC's Runs-on-request visualization is being planned, including a selection of derived variables. We are also planning development of one-way and two-way model coupling for model codes written in different languages, which will depend upon a new interactive cross-language Kamodo interface coming soon from Ensemble. This new plug-and-play approach to model coupling promises to drastically simplify the model coupling problem to a similar degree as the simplification shown in this work for direct access to model data. In addition, we are investigating how to add a HAPI interface on top of Kamodo's flythrough capability, and eventually on top of the model readers themselves. Ensemble, our development partners, is developing a system that automatically containerizes model output and enables automated container deployments on an Amazon Web Services cloud with the appropriate CloudFormation[36] scripts. In the long term, we intend to offer CCMC users these functionalities to enable reduction and visualization of model outputs on the cloud, thus reducing the data users need to download.

Maintaining this software as open-sourced code, despite the restrictions inherited from its NASA origins, is imperative to enable collaboration with Heliophysics community members – commercial, government, and academic – to improve and expand this resource. Although this capability is currently only available for a small selection of models, we are collaborating with the community to expand this resource to include additional models. We invite the reader to contact the authors to initiate such a collaboration if the reader so desires. Even now, Kamodo is growing in its range of application and utility. We are working to apply this technology to create the next generation of science software for the heliophysics and space weather communities.

---

[36] https://aws.amazon.com/cloudformation/

*Acknowledgments*

The authors acknowledge the reviewers for their guidance on significant improvements to this manuscript so far, and C. Wiegand for feedback on the manuscript. The Community Coordinated Modeling Center (CCMC, https://ccmc.gsfc.nasa.gov/) uses several Space Physics Data Facility (SPDF, https://spdf.gsfc.nasa.gov/) services, including SSCWeb and CDAWeb as mentioned in the text, to obtain input data for models and perform model-data comparisons with active and past space science missions. Work on the Kamodo-core functionality was supported by a NASA SBIR Phase 2: Space Weather R2O/O2R Technology Development grant titled "Kamodo Containerized Space Weather Models" (Contract #80NSSC20C0290) awarded to Ensemble Consultancy.
*References*

AMGeO Collaboration (2019). A Collaborative Data Science Platform for the Geospace Community: Assimilative Mapping of Geospace Observations (AMGeO) v1.0.0. Zenodo. https://doi.org/10.5281/zenodo.3564914.

Astropy Collaboration (2018). The Astropy Project: Building an Open-science Project and Status of the v2.0 Core Package. *AJ*, **156**, *3*, 123. https://doi.org/10.3847/1538-3881/aabc4f.

Bard, C., and J. Dorelli (2021). Magnetotail reconnection asymmetries in an ion-scale, Earth-like magnetosphere. *Ann. Geophys*. **39**, 991–1003. https://doi.org/10.5194/angeo-39-991-2021.

Barnum, J., A. Masson, R. H. W. Friedel, A. Roberts and B. A. Thomas (2022). Python in Heliophysics Community (PyHC): Current Status and Future Outlook. *Adv. Space Res.* Accepted. https://doi.org/10.1016/j.asr.2022.10.006.

Bilitza, D. (2018). IRI the International Standard for the Ionosphere. *Adv. Radio Sci*., **16**, 1-11, https://doi.org/10.5194/ars-16-1-2018.

Bruinsma, S (2015). The DTM-2013 thermosphere model. *J. Space Weather Space Clim.,* **5**, A1. https://doi.org/10.1051/swsc/2015001.

Codrescu, M. V., Fuller-Rowell, T. J., Munteanu, V., Minter, C. F., and Millward, G. H. (2008). Validation of the coupled thermosphere ionosphere plasmasphere electrodynamics model: CTIPe-Mass Spectrometer Incoherent Scatter temperature comparison. *Space Weather*, **6**, S09005, https://doi.org/10.1029/2007SW000364.

Cousins, E. D. P., and S. G. Shepherd (2010). A dynamical model of high-latitude convection derived from SuperDARN plasma drift measurements. *J. Geophys. Res.*, **115**, A12329, https://doi.org/10.1029/2010JA016017.

Dask Development Team (2016). Dask: Library for dynamic task scheduling. https://dask.org.

Fang, T.-W., Kubaryk, A., Goldstein, D., Li, Z., Fuller-Rowell, T., Millward, G., et al. (2022). Space weather environment during the spaceX starlink satellite loss in February 2022. *Space Weather* 20, e2022SW003193. https://doi.org/10.1029/2022SW003193.

Fok, M.-C., N. Y. Buzulukova, S.-H. Chen, A. Glocer, T. Nagai, et al. (2014). The Comprehensive Inner Magnetosphere-Ionosphere Model. *J. Geophys. Res. Space Physics*, **119**, 7522-7540, https://doi.org/10.1002/2014JA020239.

Hillger, D. W. and T. L. Schmit (2009). The GOES-13 Science Test. *Bulletin of the AMS*, **90**, 5. https://www.jstor.org/stable/26220978.

Hoyer, S. and Hamman, J. (2017). xarray: N-D labeled Arrays and Datasets in Python. *Journal of Open Research Software,* **5**, *1*, 10. DOI: https://doi.org/10.5334/jors.148.

Huba, J. and J. Krall (2013). Modeling the plasmasphere with SAMI3. *GRL*, **40**, 1, 6-10. https://doi.org/10.1029/2012GL054300.

Janhunen, P., M. Palmroth, T. Laitinen, I. Honkonen, L. Juusola et al. (2012). The GUMICS-4 global MHD magnetosphere-ionosphere coupling simulation, *J. Atmos. Solar-Terr. Phys.*, **80**, 48, https://doi.org/10.1016/j.jastp.2012.03.006.

Liu, H.-L., C. G. Bardeen, B. T. Foster, P. Lauritzen, J. Liu, et al. (2018). Development and validation of the Whole Atmosphere Community Climate Model with thermosphere and ionosphere extension (WACCM-X 2.0). *Journal of Advances in Modeling Earth Systems*, **10**, 381– 402. https://doi.org/10.1002/2017MS001232.

Kluyver, T., B. Ragan-Kelley, F. Perez, B. Granger, M. Bussonnier, et al. (2016). Jupyter Notebooks - a publishing format for reproducible computational workflows. *Positioning and Power in Academic Publishing: Players, Agents and Agendas*, 87-90, IOS Press eBooks, Clifton, VA, USA. http://dx.doi.org/10.3233/978-1-61499-649-1-87.

Maruyama, N., Sun, Y.-Y., Richards, P. G., Middlecoff, J., Fang, T.-W., Fuller-Rowell, T. J., et al. (2016). A new source of the midlatitude ionospheric peak density structure revealed by a new Ionosphere-Plasmasphere model. Geophys. Res. Lett. 43. https://doi.org/10.1002/2015GL067312.

Menzel, W. P., and J. F. W. Purdom (1994). Introducing GOES-I: The First of a New Generation of Geostationary Operational Environmental Satellites, Bulletin of the American Meteorological Society, **75**, 5, 757-782. https://doi.org/10.1175/1520-0477(1994)075<0757:IGITFO>2.0.CO;2.

Meurer, A., Smith, C.P., Paprocki, M., Čertík, O., Kirpichev, S.B., et al. (2017) SymPy: symbolic computing in Python. *PeerJ Computer Science*, 3:e103 https://doi.org/10.7717/peerj-cs.103.

Morley, S. K., Welling, D. T., Koller, J., Larsen, B. A., Henderson, M. G., et al. (2011). SpacePy - A Python-based Library of Tools for the Space Sciences. https://doi.org/10.25080/Majora-92bf1922-00c.

Pembroke, A., De Zeeuw, D., Rastaetter, L., Ringuette, R., Gerland, O., et al. (2022). Kamodo: A functional api for space weather models and data. *JOSS*, **7**, *75*, 4053, https://doi.org/10.21105/joss.04053.

Peterson, P. (2009). F2PY: a tool for connecting Fortran and Python programs. *International Journal of Computational Science and Engineering*, **4**, 4, 296–305. https://doi.org/10.1504/IJCSE.2009.029165.

Plotly Technologies Inc. (2015) Collaborative data science. *Plotly Technologies Inc.*, Montréal, QC. https://plot.ly.

Polson, S., Ringuette, R., Zheng, Y., Neihof, J., Grimes, E., and Murphy, N. (2022). Developing an Executable Paper With the Python in Heliophysics Community. *Earth and Space Science Open Archive*, https://doi.org/10.1002/essoar.10510006.1.

Qian, L., Burns, A., Emery, B., Foster, B., Lu, G. et al. (2013). The NCAR TIE-GCM: A community model of the coupled thermosphere/ionosphere system. *Geophysical Monograph Series,* **201,** 73-83, https://doi.org/10.1029/2012GM001297.

Raeder, J., Wang, Y. L., Fuller-Rowell, T. J., Singer, H. J. (2001). Global simulation of space weather effects of the Bastille Day storm. *Solar Phys.,* **204**, 325. https://doi.org/10.1023/A:1014228230714.

Rastaetter, L., R. Ringuette, D. De Zeeuw and O. Gerland (2022). Magnetic Mapping in the Inner Magnetosphere using Kamodo. Presented at the 2022 fall meeting of AGU, Dec 12-16, Chicago, IL, USA. Abstract #: SH42E-2338. https://doi.org/10.22541/essoar.167214301.16153548/v1.

Ridley, A., Deng, Y., Tóth, G. (2006). The Global Ionosphere-Thermosphere Model (GITM). *Journal of Atmospheric and Solar-Terrestrial Physics,* **68**, 839-864, https://doi.org/10.1016/j.jastp.2006.01.008.

Ringuette, R., D. De Zeeuw, L. Rastaetter, A. Pembroke, O. Gerland, and K. Garcia-Sage (2022). Kamodo's Model-Agnostic Satellite Flythrough: Lowering the Utilization Barrier for Heliophysics Model Outputs. *Frontiers in Astronomy and Space Science: Space Physics*, accepted. https://doi.org/10.3389/fspas.2022.1005977.

Robinson, R. M., L. Zanetti, B. Anderson, S. Vines, and J. Gjerloev (2021). Determination of auroral electrodynamic parameters from AMPERE field-aligned current measurements. *Space Weather*, **19**, e2020SW002677. https://doi.org/10.1029/2020SW002677.

van Rossum, G. and de Boer, J. (1991) Interactively Testing Remote Servers Using the Python Programming Language. *CWI Quarterly*, 4 (4), Amsterdam, 283–303.

Stoneback, R. A., Burrell, A. G., Klenzing, J., & Depew, M. D. (2018). PYSAT: Python Satellite Data Analysis Toolkit. *Journal of Geophysical Research: Space Physics*, 123, 5271–5283. https://doi.org/10.1029/2018JA025297.

Toth, G., De Zeeuw, D. L., Gombosi, T. I., Manchester, W. B., Ridley, A. J., et al. (2007). Sun to thermosphere simulation of the October 28-30, 2003 storm with the Space Weather Modeling Framework (SWMF). *Space Weather*, **5**, S06003. https://doi.org/10.1029/2006SW000272.

Virtanen, P., Gommers, R., Oliphant, T. E., Haberland, M., Reddy, T., et al. (2020) SciPy 1.0: Fundamental Algorithms for Scientific Computing in Python. *Nature Methods*, **17**, *3*, 261-272.

Wahr, J., S. Swenson, V. Zlotnicki, and I. Velicogna (2004). Time-variable gravity from GRACE: First results. *Geophys. Res. Lett.*, **31**, L11501, https://doi.org/10.1029/2004GL019779.

Wang, D., Y. Y. Shprits, I. S. Zhelavskaya, O. V. Agapitov, A. Y. Drozdov, and N. A. Aseev (2019). Analytical chorus wave model derived from Van Allen Probe observations. *Journal of Geophysical Research: Space Physics*, **124**, 1063. https://doi.org/10.1029/2018JA026183.

Weigel, R. S., J. Vandegriff, J. Faden, T. King, D. A. Roberts, et al. (2021). HAPI: An API Standard for Accessing Heliophysics Time Series Data. *JGR: Space Physics*, **126**, 12. https://doi.org/10.1029/2021JA029534.

Wyper, P. F., C. R. Devore, and S. K. Antiochos (2018). A Breakout Model for Solar Coronal Jets with Filaments. *ApJ*, **852**, 2. https://doi.org/10.3847/1538-4357/aa9ffc.

Zaharia, S., V. K. Jordanova, D. Welling, D., and G. Tóth (2010), Self-consistent inner magnetosphere simulation driven by a global MHD model, *J. Geophys. Res.*, **115**, A12228, https://doi.org/10.1029/2010JA015915.

Zhang, B., K. A. Sorathia, J. G. Lyon, V. G. Merkin, J. S. Garretson, and M. Wiltberger (2019). GAMERA: A Three-dimensional Finite-volume MHD Solver for Non-orthogonal Curvilinear Coordinates. *ApJS*, **244**, 20, https://doi.org/10.3847/1538-4365/ab3a4c.